\documentclass[]{aa} 
\usepackage{natbib}           
\usepackage{graphicx}         
\usepackage{amssymb}          
\usepackage{ulem}         
\usepackage[usenames,dvipsnames,svgnames]{xcolor}
\usepackage{txfonts}

\newcommand{\rmd}{{\rm d}}

\newlength{\imsize}
\newcommand{\Nabla}{\vec{\nabla}}

\newcommand{\dS}{\rmd \vec{S}}
\newcommand{\dV}{\, \rmd \mathcal{V}}
\newcommand{\surf}{{\partial \mathcal{V}}}
\renewcommand{\vec}[1]{ {\mathbf #1} }
\newcommand{\vol}{\mathcal{V}}

\newcommand{\eq}[1]{Eq.~(\ref{eq:#1})}

\newcommand{\sect}[1]{Sect.~\ref{s:#1}} 
 
\newcommand{\app}[1]{Appendix~\ref{s:#1}} 
\newcommand{\fig}[1]{Fig.~\ref{Fig:#1}}

\newcommand{\BE}{\begin{equation}}
\newcommand{\EE}{\end{equation}}
\newcommand{\BA}{\begin{eqnarray}}
\newcommand{\EA}{\end{eqnarray}}

\newcommand{\vA}{\vec{A}}

\newcommand{\vAp}{\vA_{\rm p}}

\newcommand{\vB}{\vec{B}}

\newcommand{\vBp}{\vB_{\rm p}}

\newcommand{\vBj}{\vec{B}_{\rm j}}

\newcommand{\curlA}{\Nabla \times \vA}
\newcommand{\curlAp}{\Nabla \times \vAp}  

\newcommand{\divAp}{\Nabla \cdot \vAp}

\newcommand{\divB}{\Nabla \cdot \vB}

\newcommand{\epsH}{\epsilon_{\rm H}}

\newcommand{\Ftot}{F_{\rm tot}}

\newcommand{\Hj}{H_{\rm j}}
\newcommand{\Hpj}{H_{\rm pj}}
\newcommand{\Hn}{\tilde{H}}

\begin{document} 

\title{Relative magnetic helicity as a diagnostic of solar eruptivity}
\author{
E.~Pariat\inst{1}
\and 
J.~E.~Leake\inst{2}
\and  
G.~Valori\inst{3}  
\and
M.~G.~Linton\inst{2}
\and
F.~P.~Zuccarello\inst{4}  
\and
K.~Dalmasse\inst{5}
      }
\institute{
LESIA, Observatoire de Paris, PSL Research University, CNRS, Sorbonne Universit\'es, UPMC Univ. Paris 06, Univ. Paris Diderot, Sorbonne Paris Cit\'e, 92195 Meudon, France \email{etienne.pariat@obspm.fr}
          \and
Naval Research Laboratory, Washington DC, USA 
        \and
UCL-Mullard Space Science Laboratory, Holmbury St. Mary, Dorking, Surrey, RH5 6NT, UK     
        \and
                Centre for mathematical Plasma Astrophysics, Department of Mathematics, KU Leuven, B-3001 Leuven, Belgium 
        \and
        CISL/HAO, National Center for Atmospheric Research, P.O. Box 3000, Boulder, CO 80307-3000, USA 
}

\date{Received ***; accepted ***}

   \abstract
{
The discovery of clear criteria that can deterministically describe the eruptive state of a solar active region would lead to major improvements on space weather predictions.}
{Using series of numerical simulations of the emergence of a magnetic flux rope in a magnetized coronal, leading either to eruptions or to stable configurations, we test several global scalar quantities for the ability to discriminate between the eruptive and the non-eruptive simulations.  
}
{From the magnetic field generated by the three-dimensional magnetohydrodynamical simulations, we compute and analyse the evolution of the magnetic flux, of the magnetic energy and its decomposition into potential and free energies, and of the relative magnetic helicity and its decomposition.}
{Unlike the magnetic flux and magnetic energies, magnetic helicities are able to markedly distinguish the eruptive from the non-eruptive simulations. We find that the ratio of the magnetic helicity of the current-carrying magnetic field to the total relative helicity presents the highest values for the eruptive simulations, in the pre-eruptive phase only.  We observe that the eruptive simulations do not possess the highest value of total magnetic helicity.
}
{In the framework of our numerical study, the magnetic energies and the total relative helicity do not correspond to good eruptivity proxies. Our study highlights that the ratio of magnetic helicities diagnoses very clearly the eruptive potential of our parametric simulations. Our study shows that magnetic-helicity-based quantities may be very efficient for the prediction of solar eruptions.
}

   \maketitle

\section{Introduction} \label{s:intro}




The reliable prediction of the triggering of solar eruptions is an essential step toward improving space weather forecasts. However, since the underlying mechanisms leading to the generation of solar eruptions have not yet been indisputably determined, no sufficient conditions of solar eruptivity have yet been established. Solar flare predictions are thus still strongly driven by empirical methods \citep[e.g.,][]{YuD10,Falconer11,Falconer14,Barnes16}. Nowadays most predictions rely on the determination and characterization of solar active regions through the use of several parameters and the statistical comparison with the historical eruptivity of past active regions presenting similar values of these parameters. This empirical methodology has driven the quest for the determination of which parameter, or combination of parameters, would be the best proxy of the eruptivity of solar active regions. Multiple studies have thus analyzed a relatively vast list of quantities, extracting them from existing observations of active regions and comparing them with the observed eruptivity \citep[e.g.,][]{Leka03a,Leka03b,Leka07,Schrijver07,Bobra15,Bobra16}. No simple singular parameter has yet been found to be a reliable proxy of solar eruptivity. Thanks to advances in computer science, with the development of data-mining methods and machine-learning algorithms, this search and its direct application to new flare prediction systems are now reaching a new stage of development with projects such as FLARECAST\footnote{http://flarecast.eu/}.

The determination of pertinent parameters of eruptivity has so far been almost exclusively based on observational data. Despite tremendous advances in numerical modeling of solar eruptions, few studies have used numerical simulations to advance the search for eruptivity criteria. This is partly due to the fact that numerical models are strongly driven by either having a high level of realism  \citep{Lukin11,Baumann13b,Pinto16,Carlsson16} or focus on case-by-case studies of observed events \citep{JiangC12,Inoue14,Rubiodacosta16}. Little effort is spent on performing systematic parametric simulations which would allow the determination of eruptivity criteria. Similarly to observations, the search for proxies of eruptivity requires models which are parametrically able to produce both eruptive and non-eruptive simulations. \citet{Kusano12} presents one of the few examples of such a simulation setup. By varying two parameters, \citet{Kusano12} has been able to derive an eruptivity matrix based on the relative orientation of two magnetic structures. 

Recently, \citet{Leake13b} and \citet{Leake14a}  presented three-dimensional (3D) magnetohydrodynamical (MHD) simulations of magnetic flux emergence into a stratified atmosphere with a pre-existing background coronal magnetic field. Depending on the relative orientation of, and hence the amount of reconnection between, the emerging and pre-existing magnetic flux systems, both stable and eruptive configurations were found. There have been many other studies of magnetic flux emergence, with similar parameters for the emerging flux system as those in \citet{Leake13b} and \citet{Leake14a}, and a review of some of them can be found in \citet{Cheung14}. In general, eruptive behavior requires some reconnection between emerging and pre-existing flux systems, but some previous studies also exhibit eruptive behavior without the existence of a coronal field and this external reconnection \citep[e.g.,][]{Manchester04a}. 

Along with \citet{Guennou17}, the present study is dedicated to the analysis of the eruptivity properties of the simulations of \citet{Leake13b,Leake14a}. Our goal is to determine whether or not a unique scalar quantity, computable at a given instant, is able to properly describe the eruptive potential of the system. While \citet{Guennou17} is focusing on the analysis of quantities than can be derived solely from the photospheric magnetic field, similarly to what is done with real observational data, the present work will be restricted to a few quantities computed from the coronal volume of the system.

Magnetic energy and magnetic helicity are typical scalar quantities that contribute to the description of a 3D magnetic system at any instant. Magnetic energy is known to be the key source of energy that fuels solar eruptions. However, only the so-called free magnetic energy (or non-potential energy), the fraction of the total magnetic energy corresponding to the current-carrying part of the magnetic field, can actually be converted to other forms of energy during a solar eruption \citep[e.g.,][]{Emslie12,Aschwanden14}. The non-potential energy of an active region has actually been observationally found to be positively correlated with the active region's flare index and eruptivity; it does not, however,  provide by itself a necessary criterion for eruptivity \citep{Schrijver05,JingJ09,JingJ10,Tziotziou12,YangX12,SuJ14}.

Magnetic helicity is a signed scalar quantity that quantifies the geometrical properties of a magnetic system in terms of twist and entanglement of the magnetic field lines. Magnetic helicity, together with magnetic energy and cross helicity, are the only three known conserved quantities of ideal MHD. Differently from the other two, however, magnetic helicity has the property of  being quasi-conserved even when intense non-ideal effects are developing \citep{Berger84,Pariat15b}. Magnetic helicity conservation has been suggested to be the reason behind the existence of ejection of twisted magnetized structures, that is, coronal mass ejections (CMEs), by the Sun \citep{Rust94,Low96}. There is observational evidence that magnetic helicity tends to be higher in flare-productive and CME-productive active regions compared to less productive active regions \citep{Nindos04,Labonte07,ParkSH08,ParkSH10,Tziotziou12}. Magnetic helicity conservation can also be used to predict the outcome of different magnetic interactions, for example, tunnel reconnection \citep{Linton05}. Magnetic helicity evolution has already been studied in several flux-emergence simulations \citep[e.g.,][]{Cheung05,Magara08,Moraitis14,Sturrock15,Sturrock16} but without any focus on the possible link between helicity and eruptivity. 

The goal of the present study is thus to determine whether or not quantities based on either magnetic energy or magnetic helicity are able to describe the eruptivity potential of the parametric flux emergence simulations of \citet{Leake13b} and \citet{Leake14a}, which are described in \sect{Simu}. In \sect{BandE} we present the magnetic-flux and the magnetic-energy time evolutions in the different simulations. \sect{H} analyses and compares the magnetic helicity properties of the different parametric cases. In \sect{Ecrit}, we compare the different proxies of eruptivity. We obtain the very promising results that the ratio of the current-carrying magnetic helicity to the total magnetic helicity constitutes a very clear criterion describing the eruptive potential of the simulations. In \sect{Conclusion}, we finally discuss the limitations of the present analysis and the possible application of the eruptivity criterion highlighted in this study.  

\section{Eruptive and non-eruptive parametric flux emergence simulations} \label{s:Simu}

The datasets analyzed in this study are based on the 3D visco-resistive MHD simulations presented in \citet[][hereafter L13]{Leake13b} and \citet[][hereafter L14]{Leake14a}. L13 and L14 have performed and analyzed simulations of the emergence of a twisted magnetic flux rope in a stratified solar atmosphere. The flux tube, initially located in the high-density bottom part of the simulation box (assumed to emulate the convection zone), is made buoyant and rises up in the high-temperature corona going through a minimum temperature region emulating the solar photosphere. In all seven simulations considered here, the initial conditions constituting the emerging flux rope and the thermodynamical properties of the atmosphere are kept strictly constant. The atmospheric stratification and the initial condition of the flux rope are typical of emergence simulations, and are presented in detail in L13 and L14. The simulations are also performed with the very same numerical treatment. The employed mesh is a 3D irregular Cartesian grid with $z$ corresponding to the vertical direction (the gravity direction), $y$ the initial direction of the axis of the emerging flux rope, and $x$ the third orthonormal direction. The boundary conditions are kept similar for the seven parametric simulations for which the system is evolved with the same Lagrangian-remap code Lare3D \citep{Arber01}. The numerical method is also presented in detail in L13 and L14. The simulations are performed in non-dimensional units; scaling that we are keeping in the present study. An example of a physical scaling is presented in L13 and L14.  Another choice of scaling is introduced in \citet{Guennou17}. 

The seven simulations differ by the properties of the initial ($t=0$) background field that represents the pre-existing coronal field. In these simulations an arcade magnetic field is chosen which is invariant along the $y-$direction, that is, parallel to the axis of the flux tube at $t=0$. The magnetic field is given by
\begin{eqnarray}
\vB_{Arc} &  = &  \nabla\times\vA_{Arc}, ~ \textrm{where} ,\\
\vA_{Arc} & = & B_{d}\ (0, \frac{z-z_{d}}{(x^2+(z-z_d)^2)^{3/2}}, 0)
,\end{eqnarray}
with $z_d$ the depth at which the source of the arcade is located and $B_d$ a signed quantity controlling the orientation (the sign of $B_d$) and the magnetic strength ($|B_d|$) of the arcade. This magnetic field has an asymptotic decay of $1/z^3$. The arcade is initially perpendicular to the convection-zone flux tube's axis, and thus is aligned along the same axis as the twist component of the flux tube's magnetic field. 
 The seven parametric simulations presented in this paper correspond to seven different values of $B_d$ given in Table \ref{Tab:Sim}. The simulations with a positive $B_d$, which led to a stable configuration, have been presented in L13 while the simulations with a negative $B_d$, which induced an eruption, were presented in L14. The magnetic arcade strength $|B_d|$ has  four different normalized values, that is, [0,5,7.5,10], which respectively correspond to configurations with no arcade (ND), a weak arcade (WD), a medium arcade (MD), and a strong arcade (SD), following the notation of L13 and L14.

The simulation with $B_d=0$ corresponds to the absence of coronal magnetic field. This simulation was presented both in L13 and L14 and used as a idealized reference simulation. In the real solar corona some small magnetic field is always present. L13 and L14 showed that the emergence of the flux rope in the magnetic-field-empty corona led to the formation of a stable magnetic structure in the solar corona. No eruptive behavior was observed. In the present manuscript this simulation is labeled "No Erupt ND".

\begin{figure*}
  \setlength{\imsize}{0.99\textwidth}
 \includegraphics[width=\imsize,clip=true]{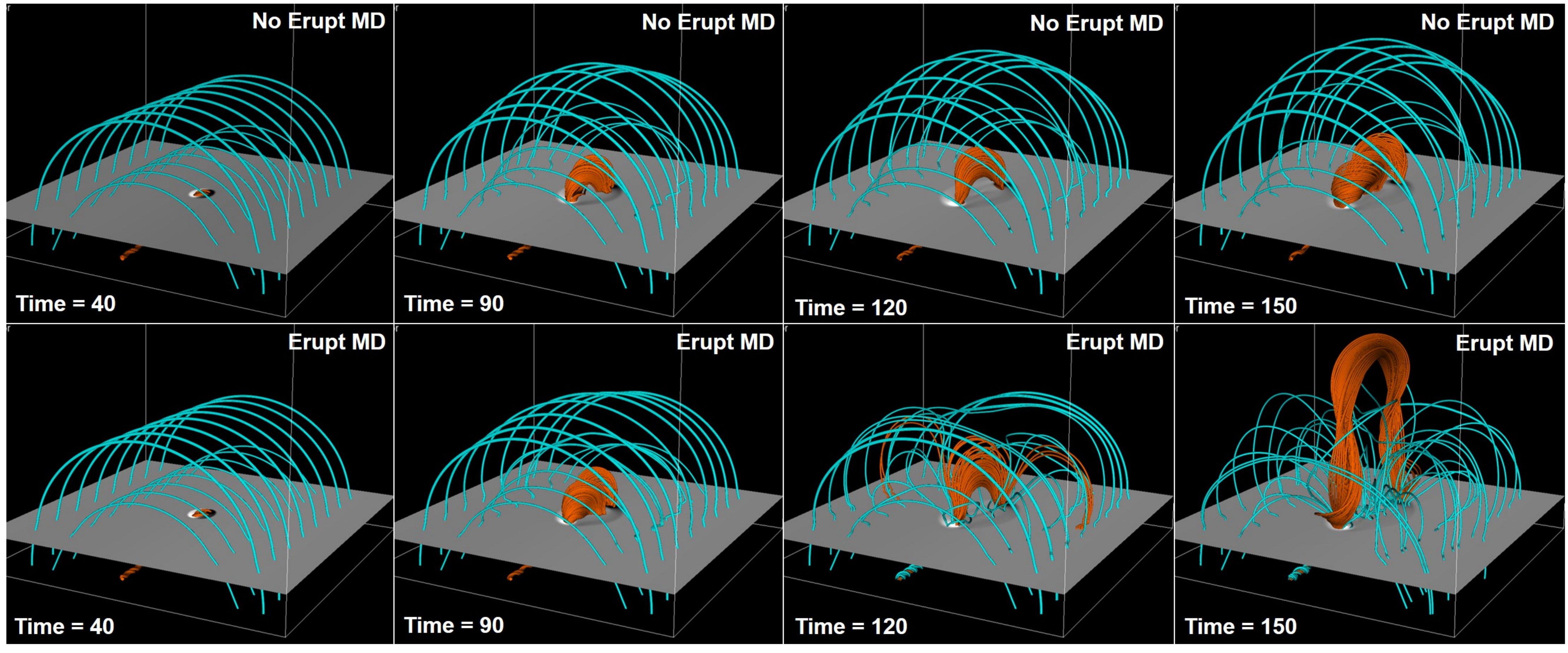}
  \caption{Snapshots comparing the evolution of the systems in the eruptive (bottom row) and non-eruptive (top row) cases with medium arcade strength ($B_d=\pm7.5)$. The respectively [cyan, orange] field lines initially belong to the [arcade, emerging flux rope]. The two-dimensional (2D) horizontal cut displays the distribution at $z=0$ of the vertical component of the magnetic field, $Bz$ with a gray-scale code. Only the volume above that boundary is considered in the present study.}
  \label{Fig:Sim}
\end{figure*}

The three simulations with $B_d>0$ were presented in L13. For this configuration the direction of the arcade field is parallel to the direction of the upper part of the poloidal field of the twisted emerging flux rope.  \fig{Sim}, top panels, presents the typical evolution of the system in the $B_d=7.5$ case. In this setup, the flux rope emerges into a field in a configuration which is unfavorable for magnetic reconnection. As the initial flux rope emerges, these simulations result in the formation of a new stable coronal flux rope. The flux rope, as with the no-arcade configuration, does not present any eruptive behavior and remains in the coronal domain until the end of the simulations. Hereafter,  these three simulations are labeled  "No Erupt WD",  "No Erupt MD", and "No Erupt SD" according to the strength of the magnetic arcade (respectively with $B_d=[5,7.5,10]$).

The three simulations with $B_d<0$ were presented in L14.  \fig{Sim}, bottom panel, displays the typical evolution of the system for the $B_d=-7.5$ simulation. For this configuration, the direction of the arcade field is antiparallel to the direction of the top of the twisted emerging flux rope. In this setup, the flux rope emerges into a coronal field for which the orientation is favorable for magnetic reconnection. Similarly to the non-eruptive simulations, the emergence results in the formation of a new coronal flux rope. However, in contrast to the $B_d\ge 0$ cases, here the new coronal flux rope is unstable and erupts. The secondary flux rope is immediately rising upward exponentially in time and eventually ejected up from the simulation domain. Once the ejection occurs, the remaining coronal magnetic field remains stable with no further eruption, even though residual flux emergence was still underway. In the following, these three simulations are labeled  "Erupt WD",  "Erupt MD", and "Erupt SD" according to the strength of the magnetic arcade (respectively with $B_d=-[5,7.5,10]$).

\begin{table*}[ht]
\caption{
Parametric simulations}
\label{Tab:Sim} 
\centering
\begin{tabular}{cccccccc}
\hline 
Label           & No Erupt SD & No Erupt MD & No Erupt WD & No Erupt ND &  Erupt WD & Erupt MD & Erupt SD \\
$B_d$           & $10$        & $7.5$       & $5$         & $0$         & $-5$      & $-7.5$   & $-10$    \\
Arcade Strength & Strong      & Medium      & Weak        & Null        & Weak      & Medium   & Strong   \\
Eruption        &  No         & No          & No          & No          & Yes       & Yes      & Yes      \\
\hline
\end{tabular}
\end{table*}

Since observational studies are not yet able to provide information about the sub-photospheric magnetic field, we only focus on analyzing the atmospheric part of the flux emergence simulations of L13 and L14. The datasets analyzed here thus correspond to the following subdomain: $x \in[-100,100]$, $y \in[-100,100]$, $z \in[0.36,150]$. To comply with the constraints of our magnetic helicity estimation routines, the original irregular 3D grid of L13 and L14 is remapped on a 3D regular and uniform mesh, using trilinear interpolation. The grid used in the present study has a constant pixel size of $0.859$ along all three dimensions. This corresponds to [1.3,2],  respectively, times the smallest horizontal and vertical grid spacing used in the original simulations in L13 and L14, located at the center of the simulation, and [0.33, 0.43] times its largest grid spacing, respectively. As discussed in \sect{Hrel}, this interpolation deteriorates the level of solenoidality ($\divB=0$) of the magnetic field compared to the original data of L13 and L14. 

The simulations last from $t=0$ to $t=200$. For all the simulations, before $t\sim 30,$ the emerging flux rope rises in the convection zone, and the coronal field (that is, the domain studied here), remains close to its initial stage. The rising flux ropes eventually reach the photospheric level, and in our datasets the emergence effectively starts around $t\sim 30$ (see \fig{BFlux}). 
In the eruptive simulations, as soon as the flux tube starts to rise in the corona, a current sheet builds up above the emerging flux rope at the boundary with the antiparallel coronal magnetic field. 
This induces continuous mild magnetic reconnection (referred to as ``external''  reconnection in Section 3.1 of L14) between the emerging flux rope and the overlying coronal field . This reconnection is weaker than the ``internal" reconnection observed from $t\sim 120$ low in the corona, within this emerging flux rope, only in the eruptive simulations of L14, Section 3.3, where reconnection within the emerging flux structure is temporally coincident with an eruption. 
The external reconnection is qualitatively characterized as `mild' relative to the internal reconnection with regards to the relative intensity of the current sheets of the reconnection flows and of the overall dynamics of the magnetic flux processed.
In the following, the period $t\in[30,120]$ will be referred to as the pre-eruptive phase both for the eruptive and the non-eruptive simulations.  An efficient proxy of eruptivity should be able to distinguish the eruptive- from the non-eruptive simulations \textit{\textbf{during}} the pre-eruptive phase only.

Intense magnetic reconnection around the flux rope enables its ejection in the eruptive simulation. The flux rope eventually crosses the top boundary of our dataset (at $z= 150$) around $t\sim 150$. The period between  $t\sim120$ and  $t\sim150$ is named hereafter as the eruptive phase. The period with $t > 150$, until the final time analyzed in this study at $t=200$, is the post-eruptive phase. It should also be noted that the non-eruptive simulations have been carried out until $t > 450$. No eruptive behavior was sighted in that later phase of their evolution. At the end of the time range studied here, the non-eruptive simulations are thus still not in an eruptive stage. Hence during the post-eruptive phase, since both the non-eruptive and the eruptive simulations are in a stable state, an efficient eruptivity proxy should not be able to discriminate between them, in addition to being able to discriminate between them in the pre-eruptive phase.

\section{Magnetic flux and energy evolution} \label{s:BandE}

Before analyzing the differences in the magnetic helicity content in the simulations, it is important to first compare the evolution of the magnetic flux and magnetic energies, quantities which are typically used to characterize eruptive systems such as active regions. More frequent and powerful CMEs are known to originate from active regions with higher magnetic flux. As discussed in \sect{intro}, they are theoretically expected and observationally found  to have a relatively high non-potential energy, and hence have a greater reserve of energy to fuel the eruption. 
\subsection{Magnetic flux}\label{s:Bflux}

\begin{figure}
  \setlength{\imsize}{0.49\textwidth}
   \includegraphics[width=\imsize,clip=true]{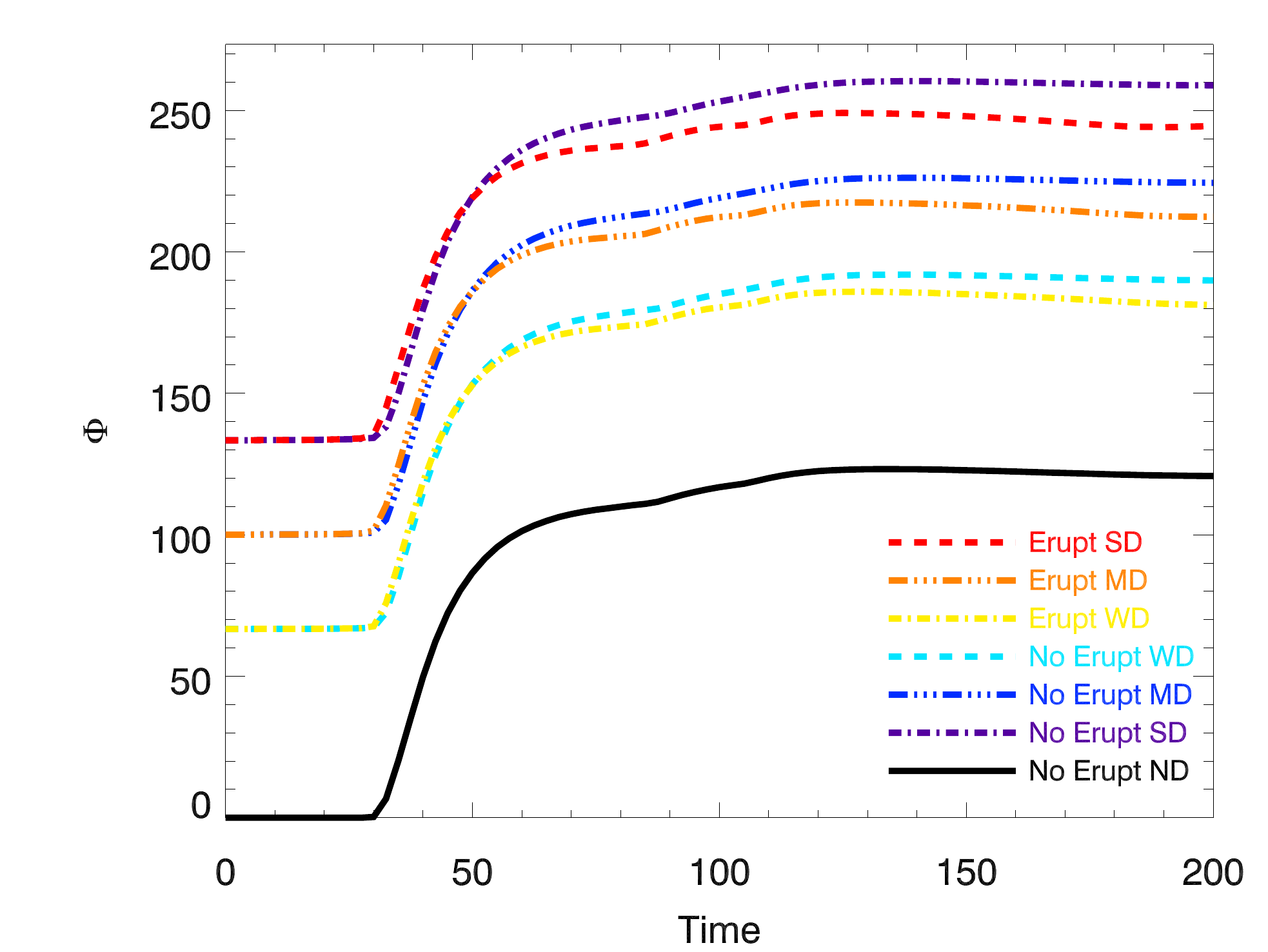}
  \includegraphics[width=\imsize,clip=true]{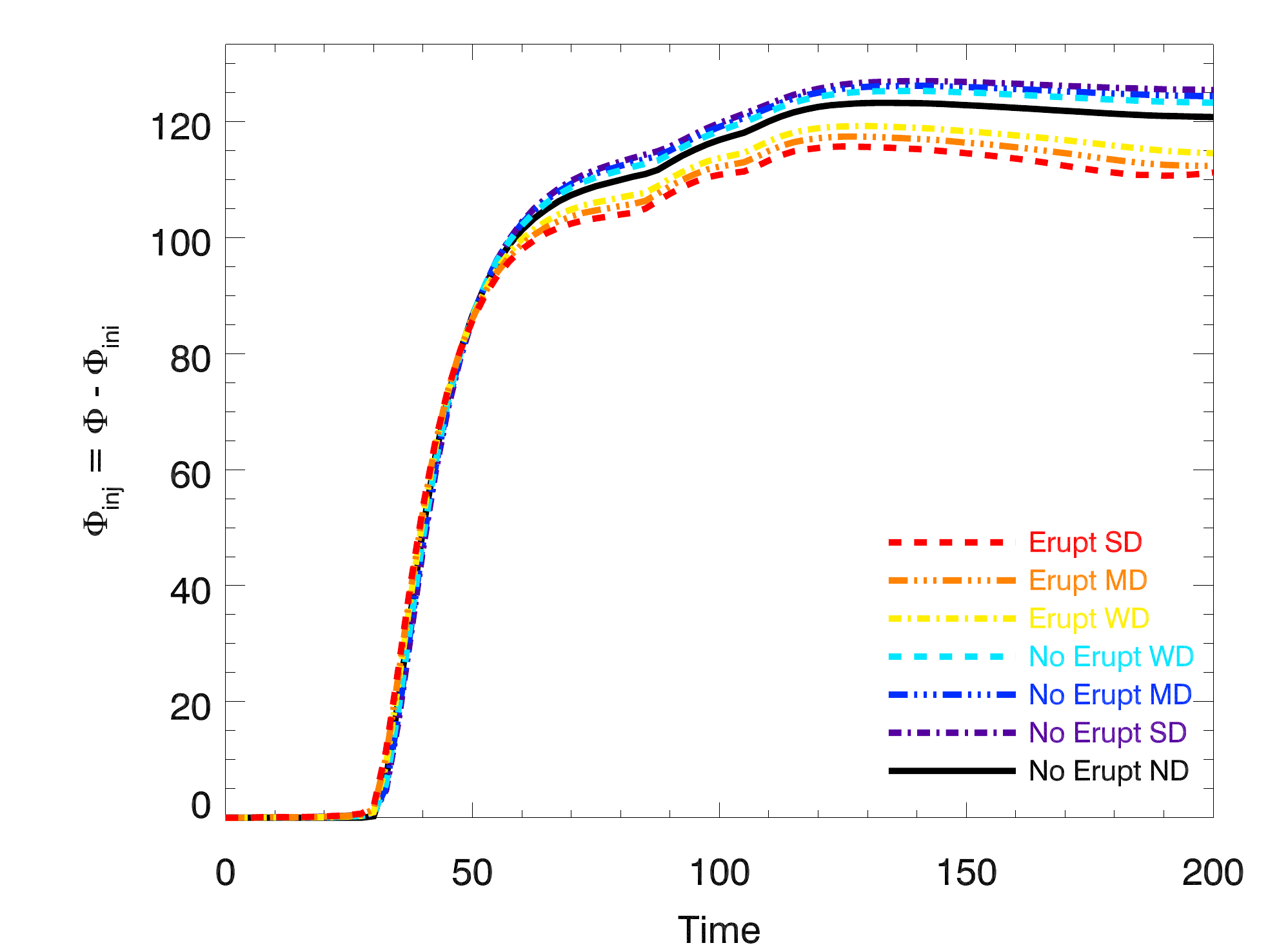}
  \caption{Evolution of the absolute magnetic flux ($\Phi$, top panel) and of the injected magnetic flux ($\Phi_{inj}\equiv\Phi-\Phi(t=0)$, bottom panel) in the system for the 7 parametric simulations. The non-eruptive simulation without surrounding field (No Erupt ND) is plotted with a continuous black line. The non-eruptive simulations with respectively [strong, medium, weak] arcade   strength, labelled [No Erupt SD, No Erupt MD, No Erupt WD], are plotted respectively with a [purple dot dashed, blue dot-dot-dot-dashed, cyan dashed] line. The eruptive simulations with respectively [strong, medium, weak] arcade   strength, labelled respectively [Erupt SD, Erupt MD, Erupt WD], are plotted respectively with a [red dashed, orange dot-dot-dot-dashed, yellow dot dashed] line.}
  \label{Fig:BFlux}
\end{figure}

The magnetic flux $\Phi$, half the total unsigned flux, at the bottom boundary of the system is given by:
\BE 
\label{eq:BFlux}
\Phi =  \frac{1}{2}\int_{z=0} |\vB \cdot \dS|  \ .
\EE
Its evolution in time for the seven different simulations is represented in the top panel of \fig{BFlux}.
Because of the different values of the strength of the arcade, the initial magnetic flux $\Phi_{ini}\equiv\Phi(t=0)$ in the seven simulations has different intensities. While the simulation with no arcade possesses no initial magnetic flux, the amount of $\Phi_{ini}$ in the other simulations is simply related to the  arcade strength. As theoretically expected, one has $\Phi_{ini, MD}/\Phi_{ini, WD} = 3/2$ and $\Phi_{ini, SD}/\Phi_{ini, MD} = 4/3$. In the present simulation framework, it is obvious that the magnetic flux does not constitute a discriminative factor for eruptivity. The values of $\Phi$ for the eruptive and non-eruptive simulations are completely intertwined. 

Because the simulation setup consists of a flux rope emerging into a coronal field, it is interesting to plot the injected magnetic flux, here defined as the flux added to the pre-existing field. In the bottom panel of \fig{BFlux}, we represent the injected magnetic flux, $\Phi_{inj}$, defined, for each simulation, in reference to its initial magnetic flux, $\Phi_{ini}$:
\BE
\label{eq:BFluxinj}
\Phi_{inj}\equiv \Phi-\Phi_{ini}  \ .
\EE
The curves of $\Phi_{inj}$ show very strong similarities in terms of injected flux. This is expected since the very same magnetic structure is emerging in all seven runs. For all seven cases, the emergence starts around $t\sim 30$ and presents a very sharp increase until $t\sim 60$. During that period, more than $80\%$ of the magnetic flux is injected in the systems. In this first phase of the emergence, the curves are barely distinguishable from one another. The curves only begin to differ slightly after $t\sim 65$.  This difference is likely due to mild external reconnections occurring in the eruptive simulations as the emerging flux rope interacts with the overlying anti-parallel field, lightly perturbing and reducing the amount of flux emerging compared to the non-eruptive runs.

After that, the magnetic injection increases moderately before reaching a plateau and slightly decreasing.  The quantity $\Phi_{inj}$ is therefore not able to discriminate the eruptive behavior present in the different simulations. No distinctive signature of eruptivity is present in the curves of the magnetic fluxes in the pre-eruptive phase.
 
\subsection{Magnetic energies}\label{s:E}

\begin{figure*}[!ht]
  \setlength{\imsize}{0.499\textwidth}
  \includegraphics[width=\imsize,clip=true]{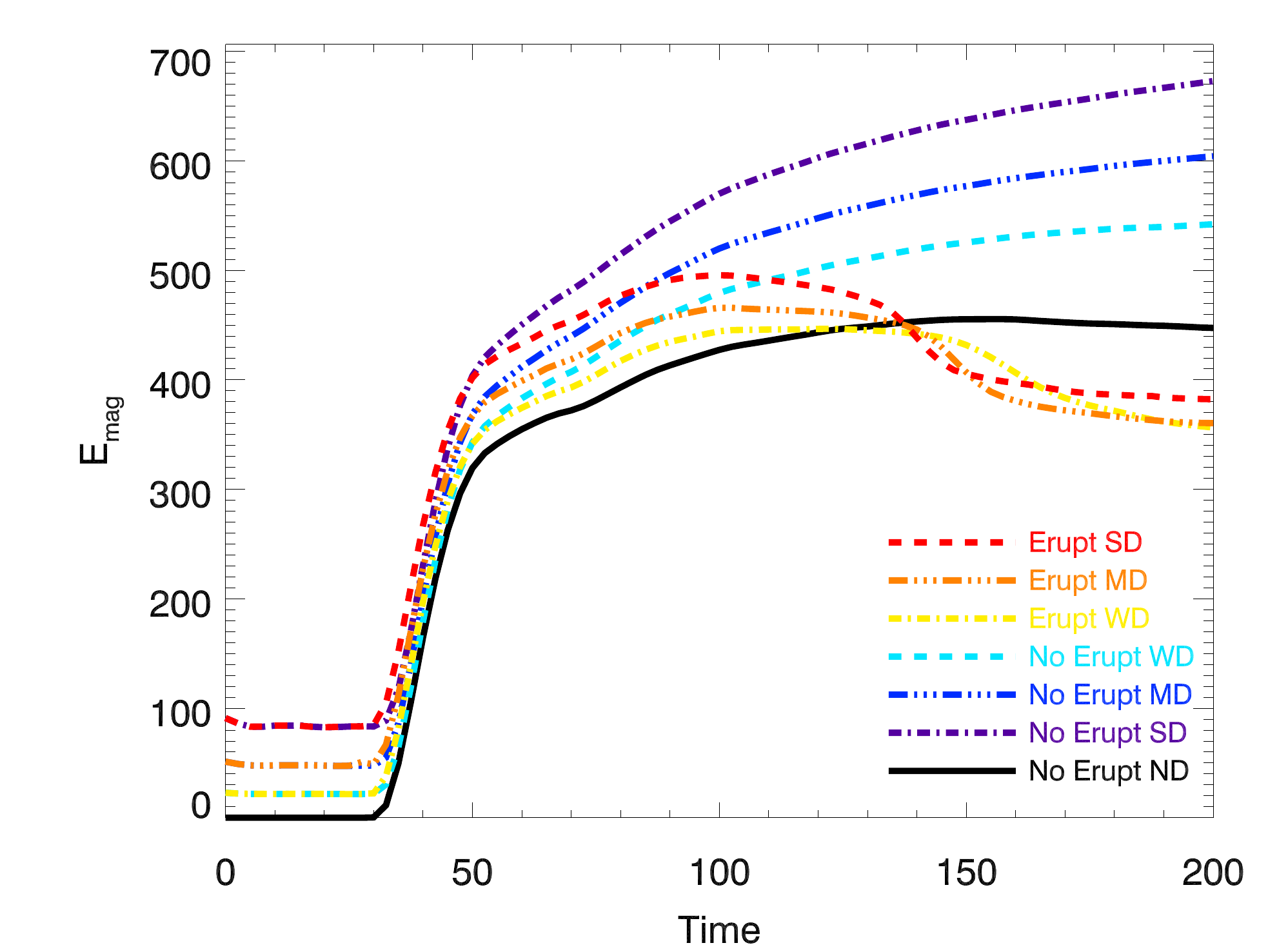}
  \includegraphics[width=\imsize,clip=true]{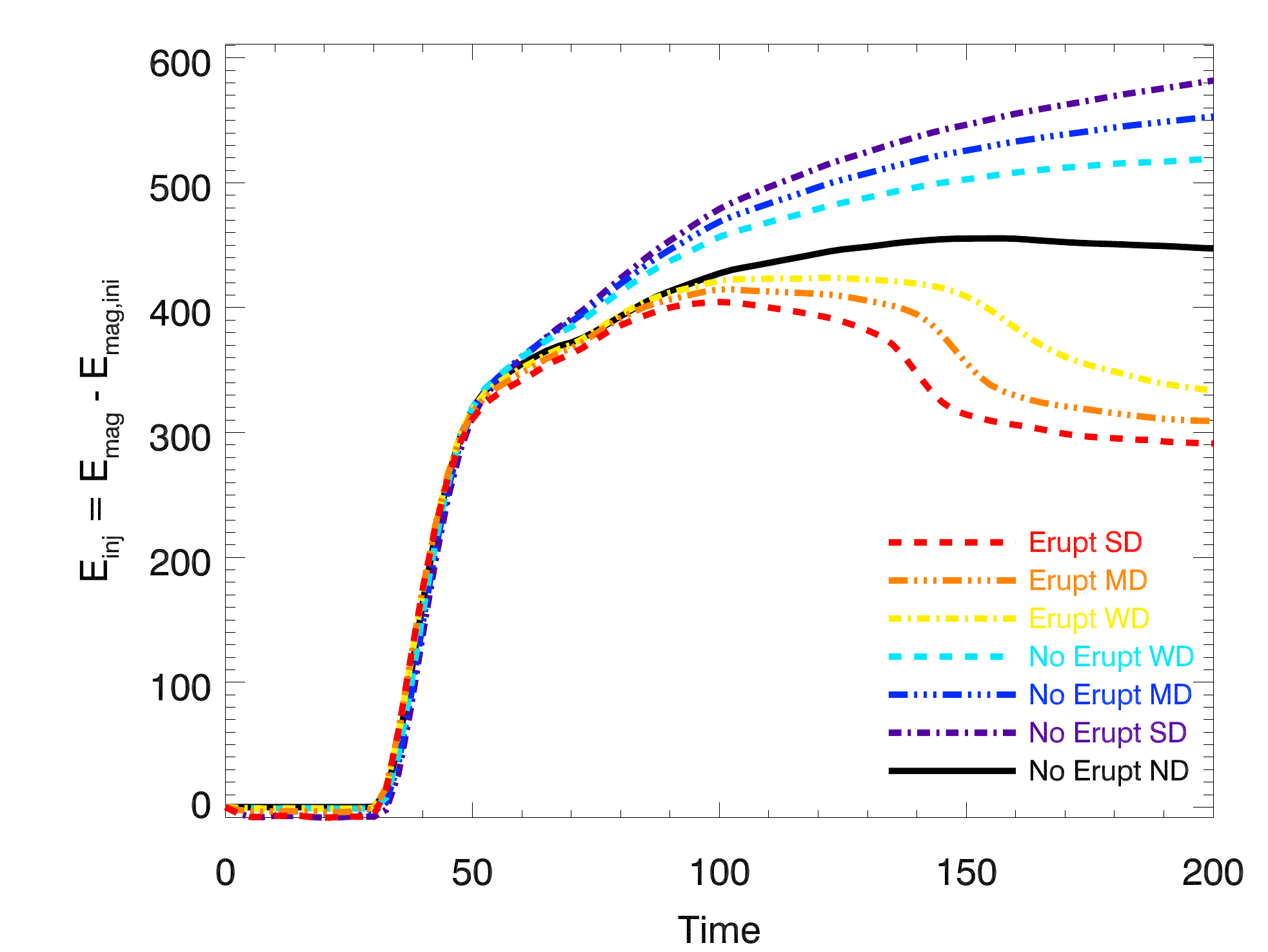}
  \includegraphics[width=\imsize,clip=true]{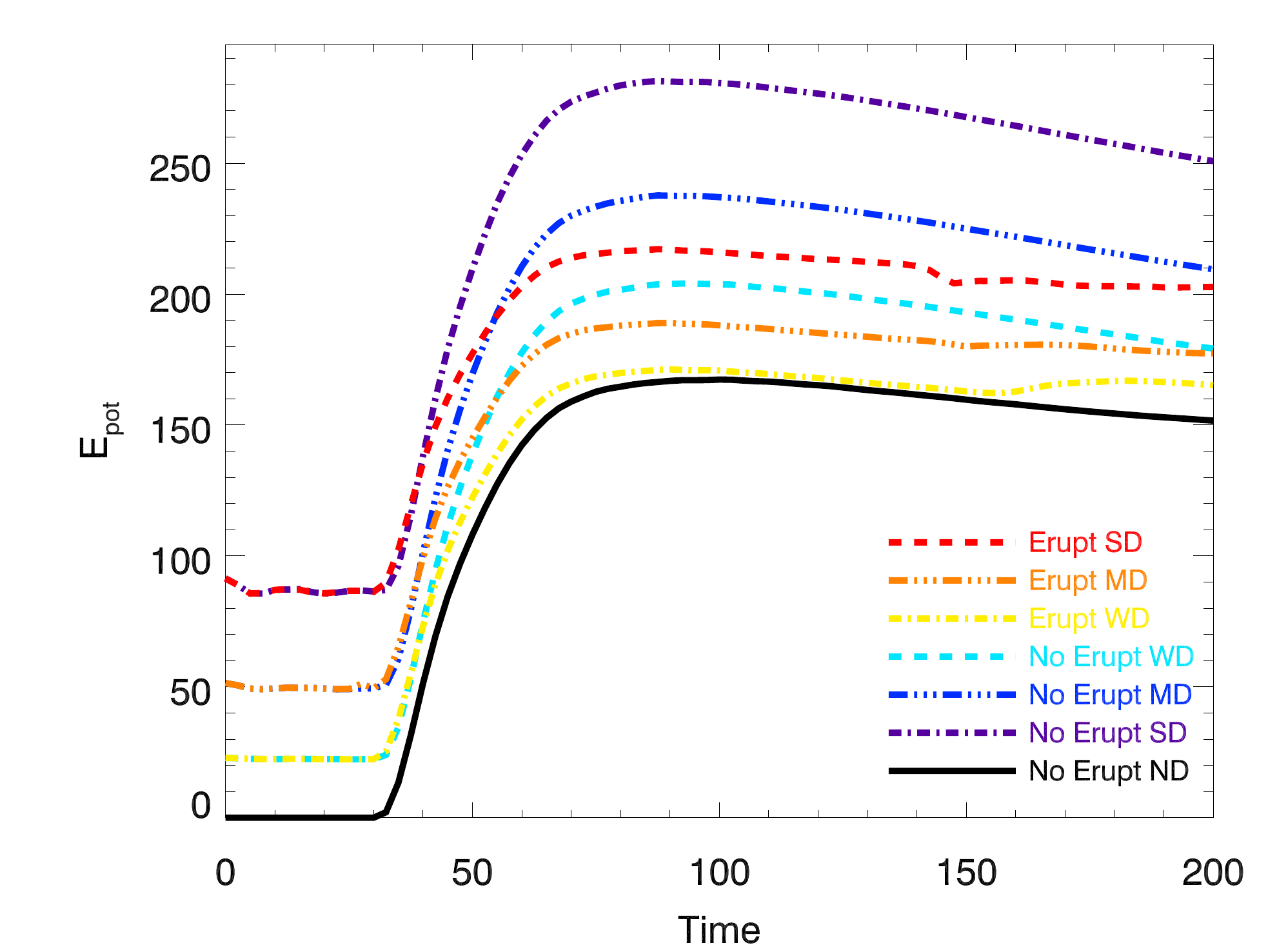}
  \includegraphics[width=\imsize,clip=true]{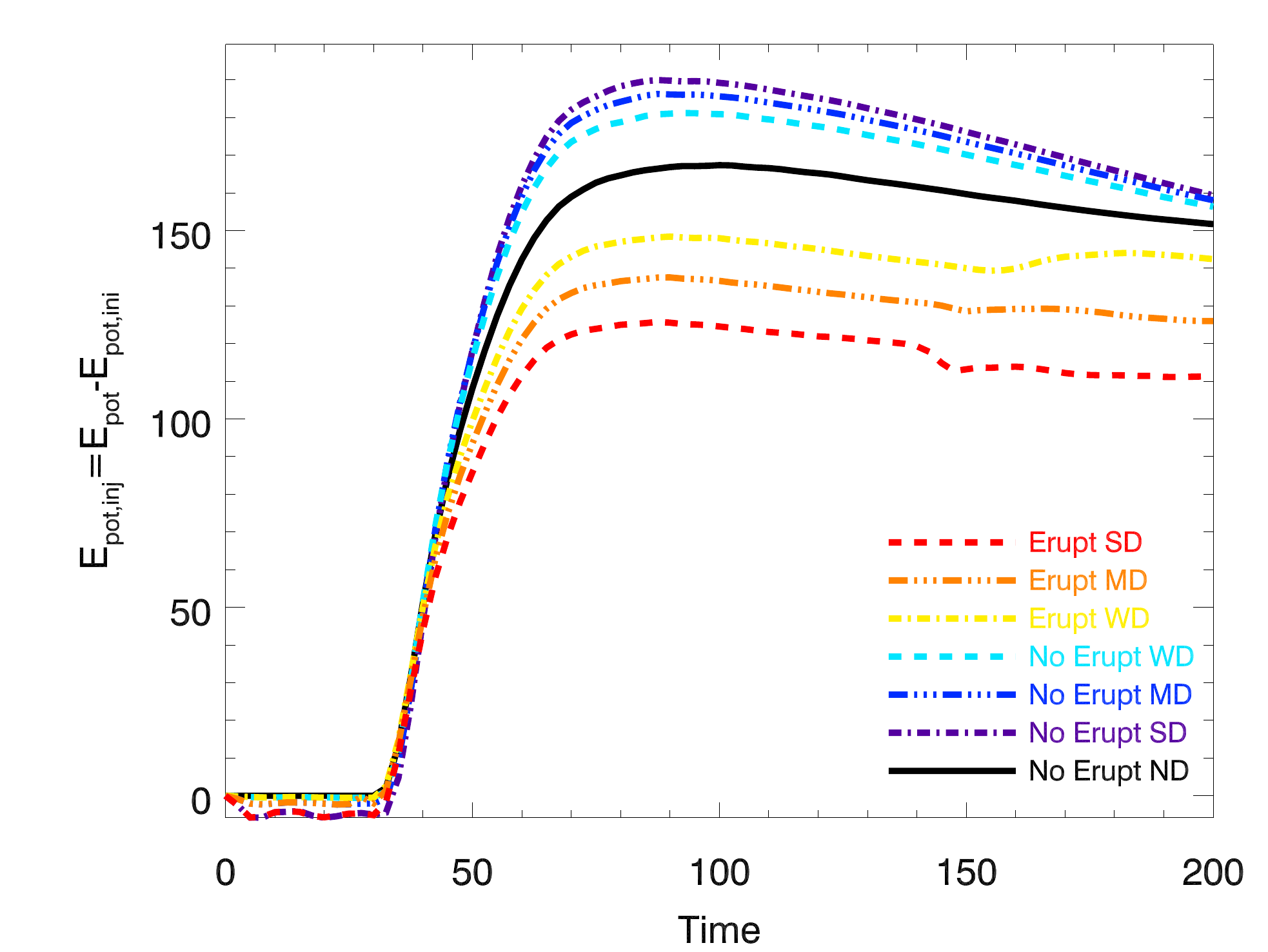}
  \includegraphics[width=\imsize,clip=true]{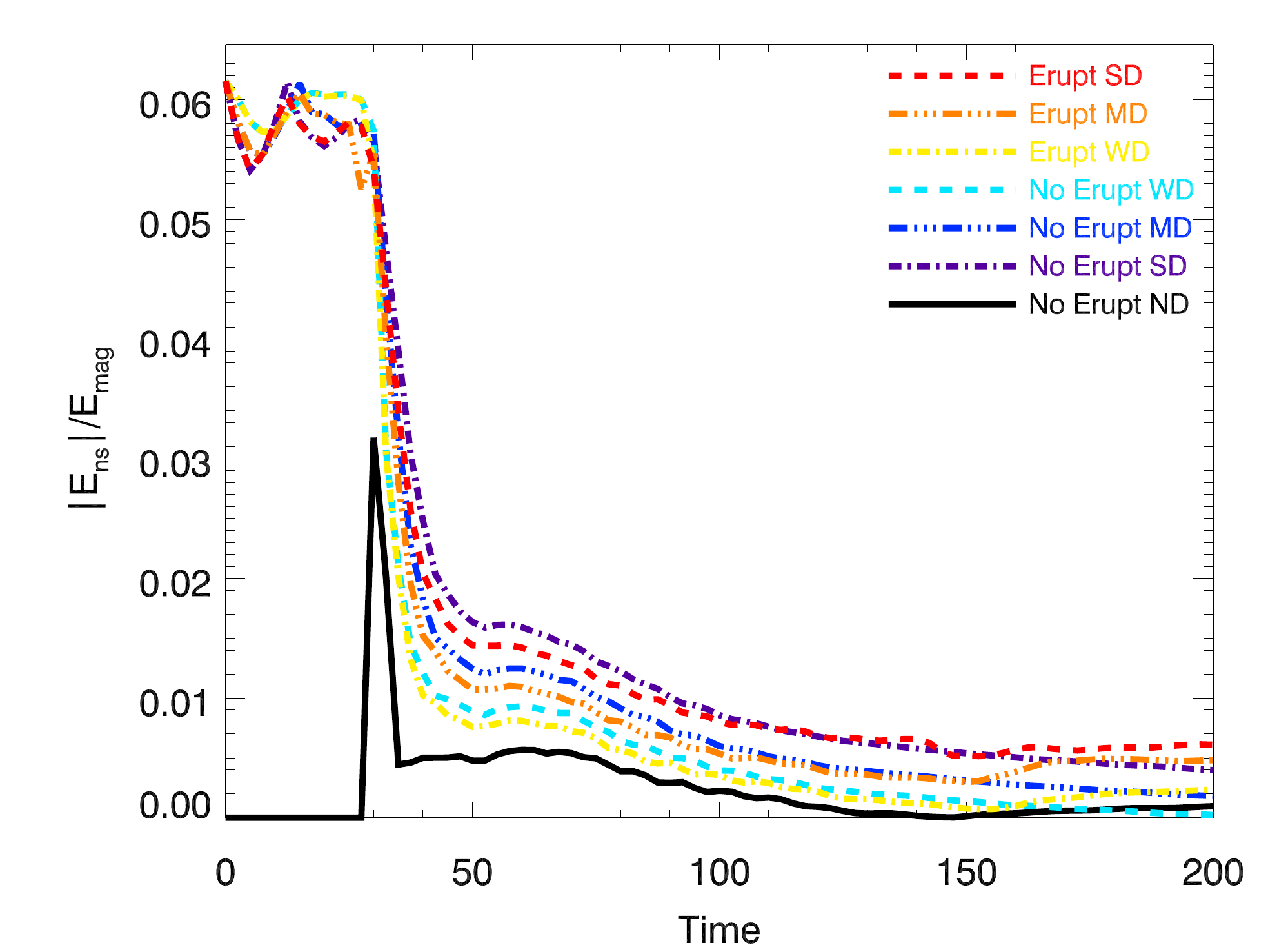}
  \includegraphics[width=\imsize,clip=true]{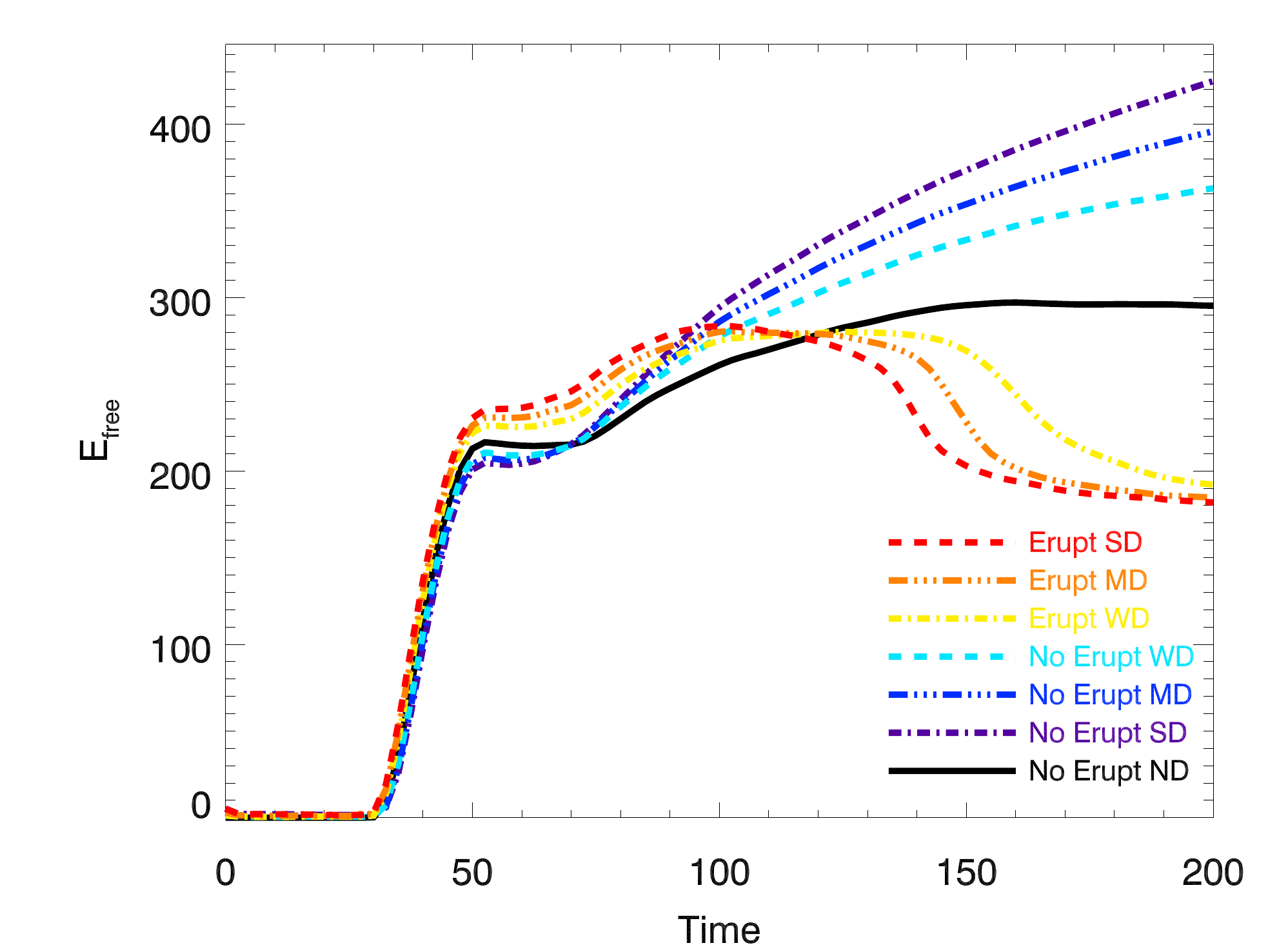}
  \caption{Energy evolution in the system for the seven parametric simulations: total magnetic energy ($E_{mag}$, top left panel), injected magnetic energy ($E_{inj}\equiv E_{mag}-E_{mag}(t=0))$, top right),  Potential magnetic energy ($E_{pot}$, middle left), potential energy variation ($E_{pot}-E_{pot}(t=0))$, middle right),  ratio of the artefact non-solenoidal energy to the total energy ($|E_{ns}|/E_{mag}$, bottom left), and free magnetic energy ($E_{free}$, bottom right). The labels are similar to \fig{BFlux}. 
}
\label{Fig:E}
\end{figure*}

The magnetic energy being the central source of energy in active solar events motivates us to present in \fig{E}, top left panel, the evolution of the magnetic energy $E_{mag}$ for the different simulations.
Similarly to $\Phi$, because of the difference in the initial magnetic coronal field, $E_{mag}$ does not constitute a pertinent criterion of eruptivity. For each simulation, we also plot the injected magnetic energy $E_{inj}$, defined relative to the initial value $E_{mag,ini}\equiv E_{mag}(t=0)$:
\BE
\label{eq:Einj}
E_{inj}\equiv E_{mag}-E_{mag,ini} \ .
\EE

As for $\Phi_{inj}$, the evolution of $E_{inj}$ for the different simulations in the initial phase of the emergence, between $t=30$ and $t=65$, presents extremely similar properties. One simulation is barely distinguishable from any other one.  In the pre-eruptive phase of the eruptive simulations, the mild external reconnection induces a lower magnetic flux injection (cf. \fig{BFlux}) that results in a slightly lower injected magnetic energy.
It is only once the system is erupting, for $t> 120$, that $E_{inj}$ starts to present significant differences between the eruptive and non-eruptive cases. This is likely due to the ejection of the erupting current-carrying structure outside of the simulation domain. In any-case, this indicates that $E_{inj}$, similarly to $\Phi_{inj}$, does not represent an efficient eruptivity criterion that would allow a forecast of the eruptions.

As discussed in \citet{Valori13}, the magnetic energy of a magnetic field with finite non-solenoidality ($\nabla\cdot\mathbf{B}\ne0$), can be decomposed as:  
\BE 
\label{eq:thomson}
E_{mag} = E_{pot} +E_{free} +E_{ns} \ ,
\EE
where $E_{pot}$ and $E_{free}$  are the energies associated with the potential and current-carrying solenoidal contributions, respectively, and $E_{ns}$ is the sum of the non-solenoidal contributions \citep[see Eqs.~(7,8) in][ for the corresponding expressions]{Valori13}. The potential field is computed such as to match the normal component of the field on all six boundaries. In the case of a purely solenoidal field, $E_{ns}=0$ in accord with the Thompson theorem. However, since numerical datasets never induce a perfectly null divergence of $\vB$, a finite value of $E_{ns}$ is generally present. Unlike other physically meaningful energies, $E_{ns}$ is a pseudo-energy quantity, which can be positive or negative.

The different values of the decomposition of energy are plotted in \fig{E}.  While the potential energy presented in the middle panel of \fig{E} does not constitute an interesting criterion for eruptivity, its evolution is interesting for understanding the energy accumulation. Initially the system is fully potential, that is, $E_{free}=0$ and $E_{mag}(t=0)=E_{pot}(t=0)$. As the flux starts to emerge, the potential energy increases due to the modification of $\vB$ at the six side-boundaries of the domain.  While at the very beginning of the emergence, for $t\in[30,40]$, the potential field of the eruptive and non-eruptive simulations shows a similar increase, the non-eruptive simulations possess a significantly higher potential energy compared to the eruptive simulation done at equivalent arcade strengths. At $t=80$, the non-eruptive simulation of a given $|B_d|$ contains approximately $1.2$ times more potential energy than its counterpart eruptive run. This further confirms that the potential energy, and its relative accumulation, cannot constitute a good eruptivity criterion.

We also notice that a large part of the injected magnetic energy, $E_{inj}$, is comprised of the increase in the potential energy, although not the majority. Before $t \sim 100$, the potential energy represents approximately one third of the accumulated total magnetic energy. This shows, as in \citet{Pariat15b}, that taking $E_{inj}$ as a proxy for the free magnetic energy $E_{free}$, as is frequently done, can lead to substantial errors, and that properly computing the energy composition of \eq{thomson}, using the full boundary information, is an important step of any proper energy budget analysis.

The free magnetic energy, $E_{free}$, is a fundamental quantity in solar eruption theory. As the primary energy tank for all the dynamics of the phenomena developing during a solar eruption, its estimation is a main focus of solar flare studies \citep{Tziotziou12,Tziotziou13,Aschwanden14}. The bottom right panel of \fig{E} presents the time variations of $E_{free}$ for the seven parametric simulations studied here. As for $\Phi_{inj}$ and $E_{inj}$, $E_{free}$ presents a relatively similar dynamic for the seven simulations in the first half of each simulation. Unlike previous criteria, the curves of $E_{free}$ for the eruptive simulations tend to be slightly higher than for the non-eruptive ones. The additional free magnetic energy remains, however, weakly higher. Interestingly, after $t\sim 100,$ the values of $E_{free}$ for the eruptive simulations tend to decrease and become lower than the ones of the non-eruptive simulations. This behavior confirms the common theoretical understanding that $E_{free}$ is indeed a quantity tightly linked with the eruptive dynamics. Nonetheless, the highest values of $E_{free}$ achieved are reached by the non-eruptive simulations in the post eruptive phase. Thus, while $E_{free}$ certainly represents a necessary condition for flares, it does not seem to be a significant sufficient condition for eruptivity. 

\begin{figure}[ht]
  \setlength{\imsize}{0.49\textwidth}
  \includegraphics[width=\imsize,clip=true]{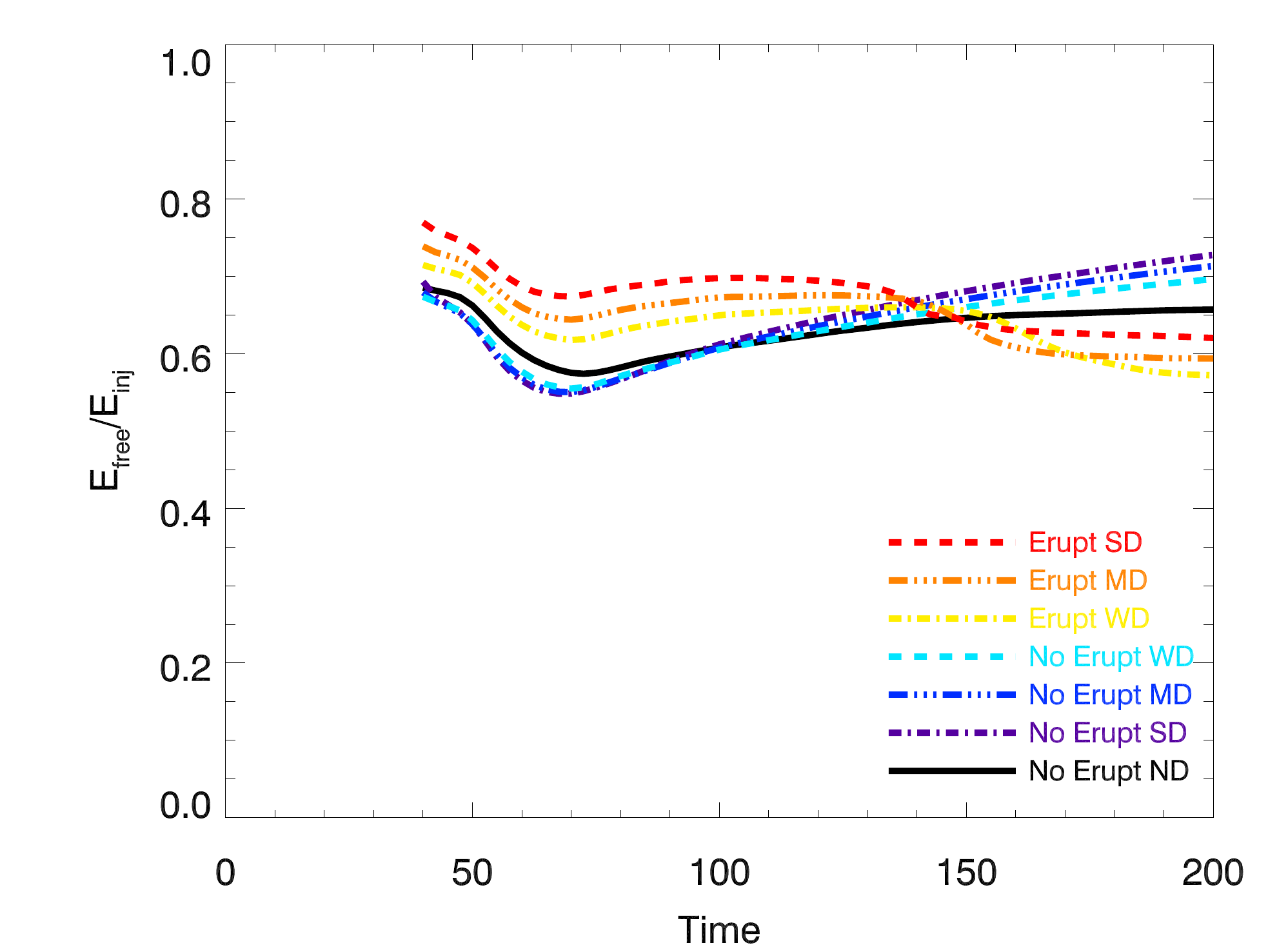}
  \caption{Time evolution of the ratio of the free magnetic energy, $E_{free}$ to the injected magnetic energy $E_{inj}\equiv E_{mag}-E_{mag}(t=0))$, for the seven parametric simulations, once the flux rope has started to emerge. The labels are similar to \fig{BFlux}. }
\label{Fig:Efreeratio}
\end{figure}

The ratio of the free magnetic energy to the injected magnetic energy, $E_{free}/E_{inj}$, is a potentially interesting proxy for eruptivity. Indeed, as shown in \fig{Efreeratio}, the eruptive simulations are well distinguishable from the non-eruptive ones already in the pre-eruptive phase using this criterion.  
For $t\in[0, 40]$, as the flux rope has not yet emerged, the injected energy is basically zero at the numerical precision. During that period, the plotted ratio is dominated by the variations on the numerical rounding precision in the numerator of the $E_{free}/E_{inj}$ ratio. We have thus disregarded this part of the evolution in the plot.
For $t\in[40, 130]$, we note that the eruptive simulations present a higher ratio of $E_{free}/E_{inj}$ than the non-eruptive simulations in the pre-eruptive and the eruptive phase. After the eruption, this quantity decreases, which is expected from a good eruptivity proxy. However, we see that the non-eruptive simulations eventually present values of $E_{free}/E_{inj}$ as high as the eruptive simulation in the initial phase. 
This evolution is related to the slight decrease of the potential energy while the free energy is still accumulating. What exactly causes the potential energy to decrease slightly at this stage is unclear. 
These simulations nonetheless do not present signs of eruptive behavior, not even beyond the time interval considered here (cf. L13). In addition, the values of $E_{free}/E_{inj}$ are, for the eruptive simulations, only slightly superior to the non-eruptive ones. In practical cases, this criterion thus may not be very efficient.

Overall, even though the free magnetic energy, and its related quantities such as  $E_{free}/E_{inj}$,  are discriminative between the eruptive and non-eruptive simulations, the difference is only marginal, in particular with regards to the criterion based on magnetic helicity that is discussed in \sect{H}. 

Following \citet{Valori13}, we also compute the ratio $|E_{ns}|/E_{mag}$ , which has been suggested as a meaningful estimation of the relative level of non-solenoidality present in the dataset. As discussed in \citet{Valori13,Valori16}, this quantity is fundamental to establishing the reliability of the magnetic helicity computation that we are presenting in \sect{H}.
 The bottom left panel of \fig{E} shows that this solenoidality criterion remains relatively small throughout the different simulations, indicating a relatively good solenoidality of the data. While for most of the simulations, $|E_{ns}|/E_{mag}\simeq 6\%$ before $t \sim 30$, the values quickly go below $2\%$ for $t > 45$. It should be noted that the levels of non-solenoidality present in the data studied in this paper are very likely much higher than the one in the original datasets studied by L13 and L14. Indeed the interpolation performed to remap the data on a uniform grid from the original staggered grid have likely significantly increased the divergence of $\vB$.  Even then, as shown in \citet{Valori16}, the low fraction of $|E_{ns}|$ presented here nonetheless ensures a good level of confidence of the magnetic helicity measurements.

\section{Magnetic helicity evolution} \label{s:H}

\subsection{Relative magnetic helicity measurements}\label{s:Hrel}

The classical magnetic helicity \citep{Elsasser56} of a magnetic field $\vB$ studied over a fixed fully-closed volume $\vol$ is gauge invariant only when considering a volume bounded by a flux surface, that is, a volume whose surface $\surf$ is tangential to $\vB$. In most practical cases, as in the present study, the studied volume surface is threaded by magnetic field. Following the seminal work of \citet{BergerField84}, we therefore track here the evolution of the relative magnetic helicity. For relative magnetic helicity to be gauge invariant, the reference field must have the same distribution of the normal component of the studied field $\vB$ along the surface. A classical choice, adopted here, is to use the potential field $\vBp$ as the reference field \citep[see][ for a possible different class of reference field]{Prior14}. As in \citet{Valori12}, we use here the definition of relative magnetic helicity  from \citet{Finn85}:
\BE 
\label{eq:Hrel}
H_V = \int_{\vol} (\vA+\vAp ) \cdot (\vB-\vBp ) \dV \ .
\EE
with  $\vA$ and $\vAp$ the vector potential of the studied and of the potential fields: $\vB=\curlA$ and  $\vBp=\curlAp$, respectively. Given the distribution of the normal component on the full surface $\vB \cdot \dS = \vBp \cdot \dS$, the potential field is unique. Independently of the time evolution of the magnetic system, this quantity is gauge invariant by definition. Even though the reference potential field may vary with time, along with possible evolution of the flux distribution on $\surf$, as demonstrated in \citet{Valori12}, it is possible and physically meaningful to compute relative magnetic helicity (called magnetic helicity hereafter) and track it in time in order to characterize the evolution of a magnetic system. 

Since magnetic helicity is an extensive quantity that scales with the square of a magnetic flux, it is of interest to study an intensive helicity-based quantity. In the following, we use the normalized helicity, $\Hn$, given by the ratio of $H_V$ to the square of the injected bottom-boundary magnetic flux, $\Phi_{inj}$ at the same time:
\BE 
\label{eq:Hnorm}
\Hn =  H_V/\Phi_{inj}^2 \ .
\EE
In the case of a uniformly twisted flux tube, the normalized helicity would correspond to the number of turns of the magnetic field. The observational properties of this normalized helicity in the solar context have been reviewed in \citet{Demoulin09}.

A possible decomposition of relative magnetic helicity from \eq{Hrel} has been given by \citet{Berger03}:
 \BA 
H_V          &=& \Hj + 2\Hpj  \quad \text{with} 
               \label{eq:HDecomp}\\
\Hj &=& \int_{\vol} (\vA - \vAp)  \cdot (\vB-\vBp) \dV 
               \label{eq:Hj}\\
\Hpj &=& \int_{\vol} \vAp \cdot (\vB-\vBp) \dV 
               \label{eq:Hpj}
,\EA
where $\Hj$ is the classical magnetic helicity of the non-potential, or current carrying, component of the magnetic field, $\vBj=\vB-\vBp$ , and $\Hpj$ is the mutual helicity between $\vBp$ and $\vBj$. The field $\vBj$ is contained within the volume $\vol$ so it is also called the closed field part of $\vB$. Because  $\vB$ and $\vBp$ have the same distribution on $\surf$, not only $H$, but also both $\Hj$ and $\Hpj$, are theoretically independently gauge invariant. An alternative and widespread decomposition splits helicity into self, potential, and mixed terms, see for example, Eqs.~(11-13) in \cite{Pariat15b}. However, since the terms in that decomposition are not individually gauge-invariant, their separate evolution is devoid of any physical meaning, and hence it is not suitable for our purpose and is not considered here.  The properties of $\Hj$ and $\Hpj$ are still poorly understood and no study has explored their dynamics, at the notable exception of \citet{Moraitis14}. 
In the numerical simulation analyzed in that work, it is worth noticing that $\Hj$ was presenting important fluctuations around the onset of the eruptions.

In order to estimate $H_V$,  $\Hj$ , and $\Hpj$ , the quantities $\vBp$, $\vAp$, and $\vA$ must be derived. The effective computation of the potential vectors requires the choice of a system of gauges. Several methods to compute $H$ in a cartesian cuboid system have been developed in recent years using different choices of gauges and/or numerical approaches. \citet{Valori16} presents a review of these methods and a benchmark of their efficiency. They found that as long as the studied field was sufficiently solenoidal, the derived helicity was overall consistent. Two of the simulations used in the present study have actually been used as test-cases of the benchmarking. 

In the present study, we adopt the method of \citet{Valori12} for the computation of the relative magnetic helicity. The potential vectors are computed using the DeVore gauge \citep{DeVore00}, that is, $A_z=0$. This method actually allows us to compute helicity with different sets of gauges (cf. \app{Ginv}) as well as allowing us to determine the quality of the helicity conservation in the numerical domain (cf. \app{Hcons}).

While $H_V$,  $\Hj$ , and $\Hpj$ are theoretically gauge invariant for purely solenoidal fields, the finite level of solenoidality of $\vB$, inherently present in any discretized dataset, induces a certain gauge dependance of the helicities \citep{Valori16}. In \sect{E}, we noted that the value of  $|E_{ns}/E|$ lies at  $<6\%$ and $<1\%$ depending on the phase of the emergence. The computation of relative helicity with different gauge sets allows us to control the impact of the finite non-solenoidality of the data on the helicity estimation. These tests are presented in \app{Ginv}. Thanks to the relatively low level of $|E_{ns}/E|$, we find that the gauge invariance of the helicity quantities is well verified, with measurement errors on the helicity quantities on the order of $5\%$. Such a tolerance can be taken as an error on the helicity curves presented here. The conclusions that are drawn from our study are not affected by such an error.

\subsection{Total magnetic helicity evolution comparison} \label{s:Hevol}

\begin{figure}[ht]
  \setlength{\imsize}{0.49\textwidth}
   \includegraphics[width=\imsize,clip=true]{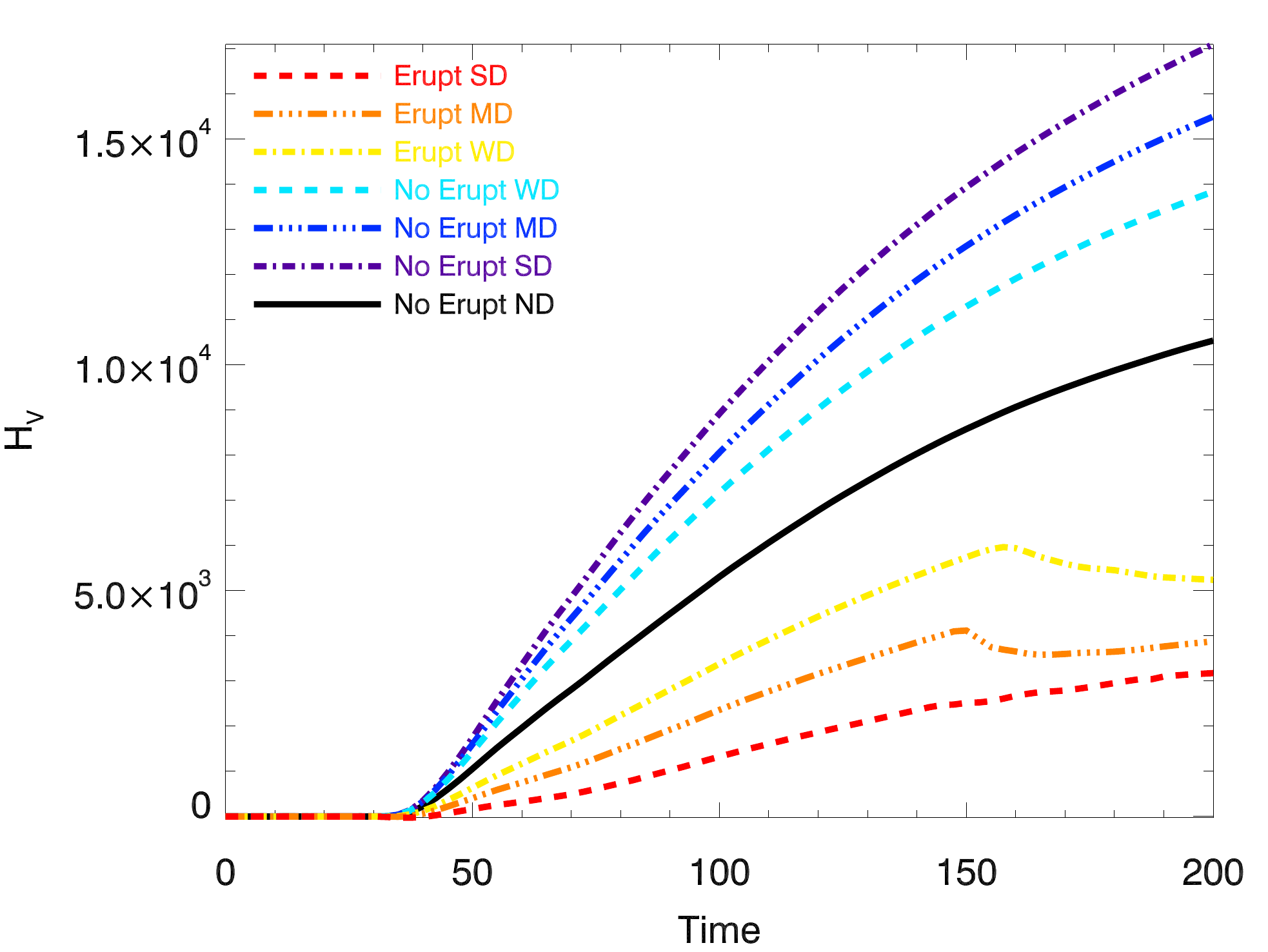}
  \includegraphics[width=\imsize,clip=true]{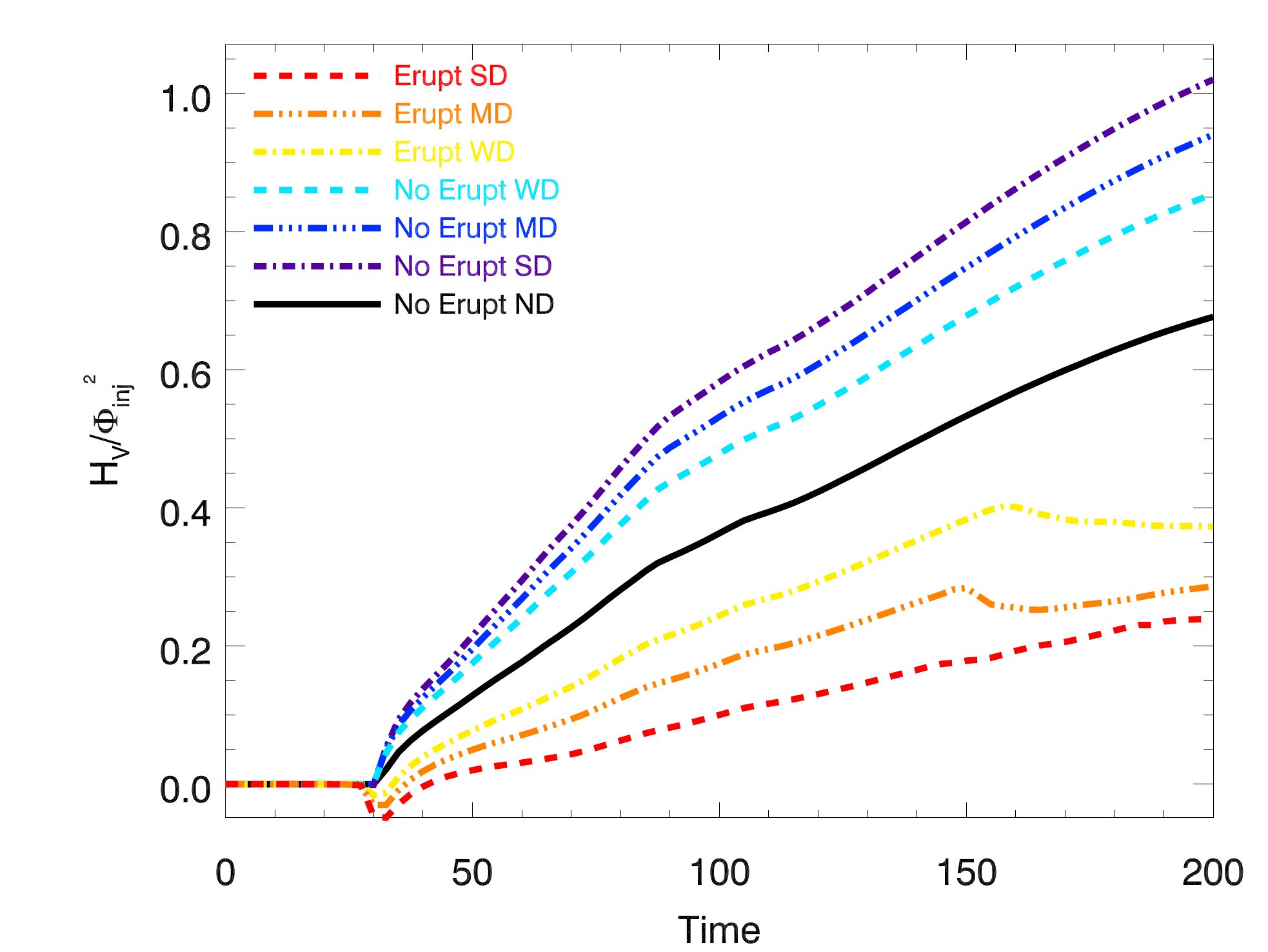}
  \caption{$H_v$ (top panel) and $\Hn$ (bottom panel) evolution for the seven parametric simulations. The labels are similar to \fig{BFlux}. 
}
  \label{Fig:Hevol}
\end{figure}

The comparative evolution of magnetic helicity for the seven different flux emergence simulations is presented in \fig{Hevol}. In all cases, the helicity presents a smooth increase in time as helicity is injected into the system thanks to the continuous emergence. In the eruptive cases, the ejection of the flux rope from the simulation domain is associated (very weakly in the strong  arcade case) with a small decrease of $H_V$ after $t \sim 150$. It is worth noting that, compared to the injection of magnetic flux and energy, the helicity accumulates much more smoothly and slowly. While more than $70\%$ of $\Phi_{inj}$ and $50\%$ of $E_{free}$ is injected into the system between $t=30$ and $t=50$, only $10\%$ of $H_V$ has been injected into the system during that period. In all the flux emergence simulations studied here, the helicity injection is thus partly delayed compared to these other quantities. This delay between the magnetic flux increase and the helicity accumulation agrees with the trend noted in observational  studies of active region emergence \citep{Jeong07,TianL08,LiuY12}.

The first significant result is that, unlike for magnetic flux and magnetic energy, magnetic helicity presents very significant differences between the different simulations immediately after the very start of the emergence. Each simulation is easily distinguishable from the others as early as $t \sim 30$.  The magnetic helicity is much more of a discriminant than the energies and accumulated magnetic flux are. Magnetic helicity is thus able to characterize very well the magnetic configuration, as it depends not only on the strength of the surrounding field, but also on its orientation relative to the emerging flux rope.

As discussed in the Introduction, a large absolute value of the total magnetic helicity has been frequently suggested as a potential proxy for flare eruptivity. In the framework of the present simulation, however, we notice that this is not the case.  The top panel of \fig{Hevol} shows that the non-eruptive simulations all have a total absolute helicity $|H_V|$ several times higher than the eruptive one. Similarly, the normalized helicity (\fig{Hevol}, bottom panel) presents higher values for the non-eruptive cases. Our results indicate that a large value of $|H_V|$ cannot be used as a criterion for eruptivity. 

Looking at the influence of the arcade field strength, we note that there is an opposite behavior of the eruptive and non-eruptive simulations when it comes to the total helicity. For the non-eruptive simulations, the stronger the arcade field, the greater the total helicity $|H_V|$ (and $\Hn$), while for the eruptive cases, the strength of the arcade and the intensity of $|H_V|$ are anti-correlated.

The origin of these behaviors can be first explained by the fact that the weaker the  arcade strength, the closer the system is to the no-arcade case (an infinitely weak arcade   would effectively correspond to an absence of  arcade field). This explains why, given the orientation of the arcade, the curves of $H_V$ and $\Hn$ tend to converge to the no-arcade case as the arcade strength decreases. The curves of $\Hn$ and $H_V$ thus tend to become lower or higher, respectively, for the non-eruptive and eruptive cases, as the arcade strength becomes lower.  

This however does not explain why the orientation of the arcade leads to a higher $H_V$ in the non-eruptive case and lower one for the eruptive simulations. This dependence originates from the fact that, unlike most quantities, magnetic helicity is intrinsically non-local \citep{Berger00}. When the flux rope is emerging, it not only advects its own helicity, but also instantaneously exchanges helicity with the surrounding magnetic field. As we show in the following simplified toy model, the difference of $H_V$ between the different simulations is directly marked by the mutual helicity shared by the emerging flux rope and the arcade field.

In the case of a system formed by two closed flux tubes, the total helicity is the sum of the proper helicity contained in each of the flux tubes, their self helicity, plus the helicity shared between the flux tubes, their mutual helicity \citep[cf. e.g.][]{Berger84,Berger03}. In an analogous toy model, one can theoretically decompose the helicity of the present system between the self helicity of the emerging flux rope, $H_{S,FR}$, the self helicity of the arcade field, $H_{S,Arc}$, and the mutual helicity shared between the emerging flux rope and the arcade field, $H_{mut}$:
 \BE 
H_V  =H_{S,Arc} +  H_{S,FR} + H_{mut}\ .
\EE
 
Initially, the flux rope not having yet emerged, the helicity of the system is solely given by the helicity of the arcade field and is expected to be null since the system is initially quasi-potential in the coronal domain (cf. \fig{E}), independently of the strength of the arcade field: $H_V(t=0) = H_{S,Arc} \simeq 0$. This is confirmed by the measured values of helicity at the beginning of the simulations in all cases. Furthermore, the simulation with no-arcade field contains no mutual helicity.  The values of $H_V$ in that case should roughly represent the evolution of the self helicity of the flux rope field: that is, $ H_{S,FR} (t) \sim H_{V, \textrm{ No Erupt ND}}(t)$.

The differences in the curves of $H_V$ should therefore originate from the difference in $H_{mut}$. In the case of two closed curved flux ropes, their mutual helicity is equal to the product of their magnetic flux weighted by their Gauss linking number \citep{BergerField84,Berger00}. Depending on the relative orientation of the curves, the linking number can either be positive or negative. In the present simulation, it is reasonable to argue that $|H_{mut}|$ will be proportional to the product of the flux of the magnetic arcade, $\Phi_{Arc}$, and the flux of the magnetic flux rope $\Phi_{FR}$. The flux of the emerging flux rope is roughly constant between the simulations, and almost exactly in the initial phase of the emergence, before $t\sim65$ as shown by the evolution of $\Phi_{inj}$ in \fig{BFlux}. For each simulation, the flux of the arcade is directly given by the values of the flux initially, that is, $\Phi_{Arc}=\Phi_{ini}$. The sign of the mutual helicity depends on the relative orientation of the arcade and the emerging flux rope. When the arcade and the axial field of the flux rope have a positive crossing, for the non-eruptive simulation, they should have a positive mutual helicity, while they should present a negative mutual helicity when the magnetic field orientation between the two has a negative crossing, for the eruptive cases. The helicity in the system should thus follow the relation:
 \BE 
H_V  -  H_{V, \textrm{ No Erupt ND}} \equiv H_D \sim  H_{mut} \sim \pm \xi \Phi_{Arc}  \label{eq:Toymodel} 
,\EE
with $\xi$ a constant of proportionality, and where the plus or minus sign applies to the non-eruptive or eruptive cases, respectively.

Qualitatively, this toy model predicts that the eruptive simulation should have a lower $H_V$ than the no arcade case while the non-eruptive one should have greater values. In addition, the stronger the arcade field, the further away $H_V$ is from the no-arcade case. Quantitatively one also finds very good agreement between the \eq{Toymodel} predicted by this simple toy model and the measured values of $H_V$ . During the main part of the emergence, when the system is not too affected by the ejection of magnetic field, one measures at $t=75$ that $H_{D, \textrm{No Erupt MD}}/H_{D, \textrm{No Erupt WD}}= 1.46$, and $H_{D, \textrm{Erupt MD}}/H_{D, \textrm{Erupt WD}}= 1.51$, while our toy model theoretically predicts that these ratios should be equal to the ratio of the arcade strength between the medium arcade and the weak arcade case, that is, $\Phi_{ini, MD}/\Phi_{ini, WD} = 3/2$. Similarly, one has $H_{D, \textrm{No Erupt SD}}/H_{D, \textrm{No Erupt MD}}= 1.30$, and $H_{D, \textrm{Erupt SD}}/H_{D, \textrm{Erupt MD}}= 1.3$, which should be equal to $\Phi_{ini, SD}/\Phi_{ini, MD}= 4/3$ according to our toy model. In the case of the non-eruptive simulations, the agreement improves as the emergence further develops. 

The excellent agreement between these values demonstrates the importance of the mutual helicity between the emerging structure and the surrounding field, which, being added to or subtracted from the helicity advected by the emerging flux rope, significantly modifies the total amount of helicity. Even though the emerging structure is exactly the same in the seven simulations, the helicity budget is profoundly modified by the surrounding field. This highlights the importance of the surrounding environment when considering the budget of magnetic helicity in flux emergence regions, unlike with more classical quantities such as energies.

While self and mutual helicity are useful theoretical concepts, they are in practice very difficult to use. The distinction between the emerging field and the arcade field is strongly subjective. When only considering a unique snapshot of  one of the simulations, it is very difficult to objectively disentangle these two structures. It is even more difficult to directly compute each contribution. It is only thanks to the combined seven parametric simulations that we are able to compute the respective self and mutual contributions in the present study. Unlike $\Hj$ and $\Hpj$, we do not believe that it is generally possible to estimate these quantities from general datasets.

\subsection{Current-carrying magnetic helicity evolution comparison} \label{s:Hjevol}

We have seen that while the total magnetic helicity $H_V$ is very discriminative of the different parametric simulations, its use to predict the eruptive behavior is limited, since for a given magnetic flux injection, the eruptive simulations posses a lower total amount of helicity. We will now discuss the evolution of the terms $\Hj$ and $\Hpj$ decomposing the magnetic helicity (cf. \eq{HDecomp}) and  show that they constitute a very promising criterion of eruptivity.

\begin{figure}[ht]
  \setlength{\imsize}{0.49\textwidth}
  \includegraphics[width=\imsize,clip=true]{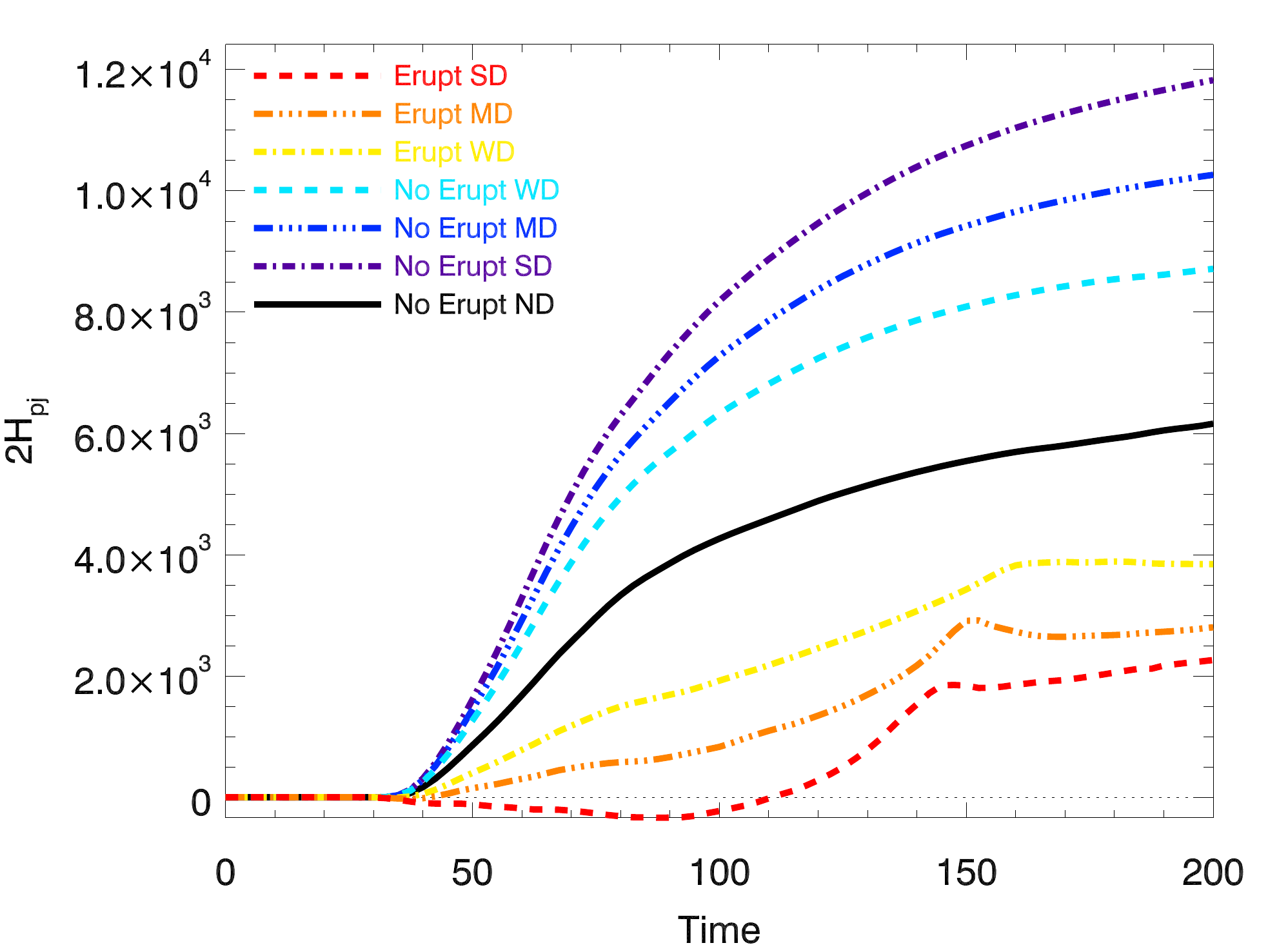}
  \includegraphics[width=\imsize,clip=true]{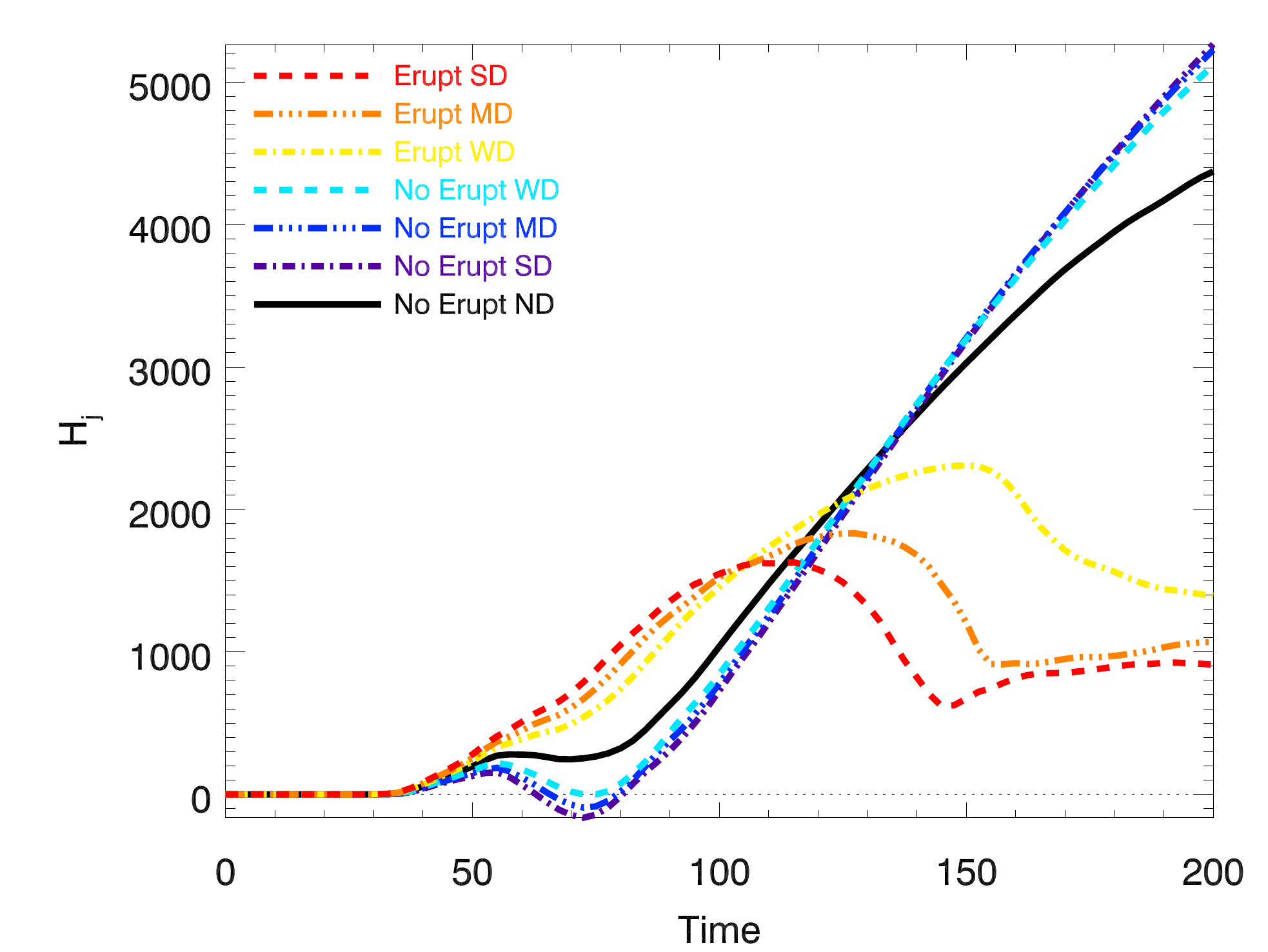}
  \caption{Time evolution of $2\Hpj$ (top panel), and $\Hj$ (bottom panel) for the seven parametric simulations. The labels are similar to \fig{BFlux}
}
  \label{Fig:HjHpj}
\end{figure}

The time evolution of $\Hj$ and $\Hpj$ for the seven parametric simulations is presented in \fig{HjHpj}. We note that while the curves of $\Hpj$ present a relative distribution very similar to $H_v$ between the different simulations, the curves of $\Hj$ differ significantly. Regarding $\Hpj$, the non-eruptive simulations have an evolution similar to one another and similar to $H_V$ both in shape and in amplitude. The differences in the evolution of $\Hpj$ for the eruptive simulations is more marked. While the weaker-arcade-strength case presents a relatively smooth increase, the stronger-arcade case displays slightly negative values during the first part of the emergence until $t\sim 100$, and then some increase. As for $H_V$, no specific behavior before the onset of the eruption is noticeable in the evolution of $\Hpj$.

The second main result of our study is that the curves of $\Hj$ are, on the contrary, strongly marked by the eruptive behavior (\fig{HjHpj}, bottom panel). We note that the non-eruptive and eruptive simulations present two very distinct groups. The simulation without surrounding magnetic field separates the two groups of simulations. For that simulation, after a slight increase for $t\in[30,55]$, $\Hj$ presents a plateau until $t\sim75$, before presenting a slow and steady increase. The three non-eruptive simulations with a surrounding magnetic field present a relatively similar behavior to one another. They are tightly grouped and are similar to the no arcade case. Instead of presenting a plateau  between $t=30$ and $t=50$,  $\Hj$ decreases, even reaching negative values before steadily increasing after $t\sim 75$.  These negative values may be related to the opposite direction of the current carrying field, mostly dominated by the emerging flux rope, and the vector potential of the potential field, mostly dominated by the arcade field.
 Unlike for $H_V$ and $\Hpj$, the field strength of the arcade does not seem to significantly influence the evolution of the values of $\Hj$, although we note that the weak arcade curve is the one closest to the no arcade case.  This is probably because $\Hj$, being related to the current carrying field, is mostly influenced by the emerging flux rope rather than the initially potential arcade.

The eruptive simulations are tightly grouped with one another. Unlike the non-eruptive simulations, they present a quasi-steady increase in the first half of the simulation, before $t \sim 100$. The curves then eventually reach a maximum, decrease, and then remain relatively constant. While in the first part of the simulation, the curves only differ slightly in intensity, the timing of the maximum and the subsequent evolution is strongly influenced by the arcade field strength. The occurrence time of the maximum is anti-correlated with the strength of the arcade. The stronger the  arcade, the earlier the peak of $\Hj$. This is likely correlated to the difference in the eruption time for the different eruptive simulations. As noted in Figure 12 of L14, the stronger the  arcade, the earlier the flux rope moves and is eventually ejected, leaving the domain. The differences between the curves of $\Hj$ for the eruptive simulations have the following explanation: the stronger the external arcade, the larger is the flux available for reconnection. For the emerging flux rope to erupt, the shell of stabilizing field surrounding it must be removed. Since the emergence timescale is dictated by the same photospheric evolution, more flux is available for reconnection for a given flux-rope emergence rate, the peeling of the outer shell  is faster , and the time of eruption is 
earlier . Hence, the stronger the external dipole field, the earlier the start of the eruption. As a corollary, the longer it takes to erupt, the more flux rope emerges, therefore, the higher is the maximum of $H_j$ that can be reached.

While the eruptive simulations all display a higher value of $\Hj$ in the initial phase of the flux emergence, before $t \sim 100$, overall, the non-eruptive simulations are the ones that present the highest values of $\Hj$ in the later time of the evolution. Hence, the value of $\Hj$ alone, while clearly being  affected by the eruptive behavior, cannot directly be used as an eruptivity criterion. The high value of $\Hj$ in the second phase of the simulations would otherwise suggest that the non-eruptive simulations could become unstable, which does not agree with the dynamics observed at the end of these simulations. The situation is somehow similar to what was found for the free magnetic energy (cf. \fig{E}, bottom right panel), with the difference being that helicity spreads the curves farther apart, that is, discriminates better between the different cases.

If $\Hj$ itself does not constitute an obvious eruptivity criterion, it nonetheless represents a significant portion of the total helicity of the system for the eruptive simulations, as can be seen in \fig{HjHpjgauges} for the medium arcade case,  for example. Actually, the fraction of $\Hj$ to the total helicity is a key distinction between, first, the eruptive and, second, the non-eruptive simulation, but also between the pre-eruptive phase and the post-eruptive phase of the eruptive simulations.

\begin{figure}[ht]
  \setlength{\imsize}{0.49\textwidth}
  \includegraphics[width=\imsize,clip=true]{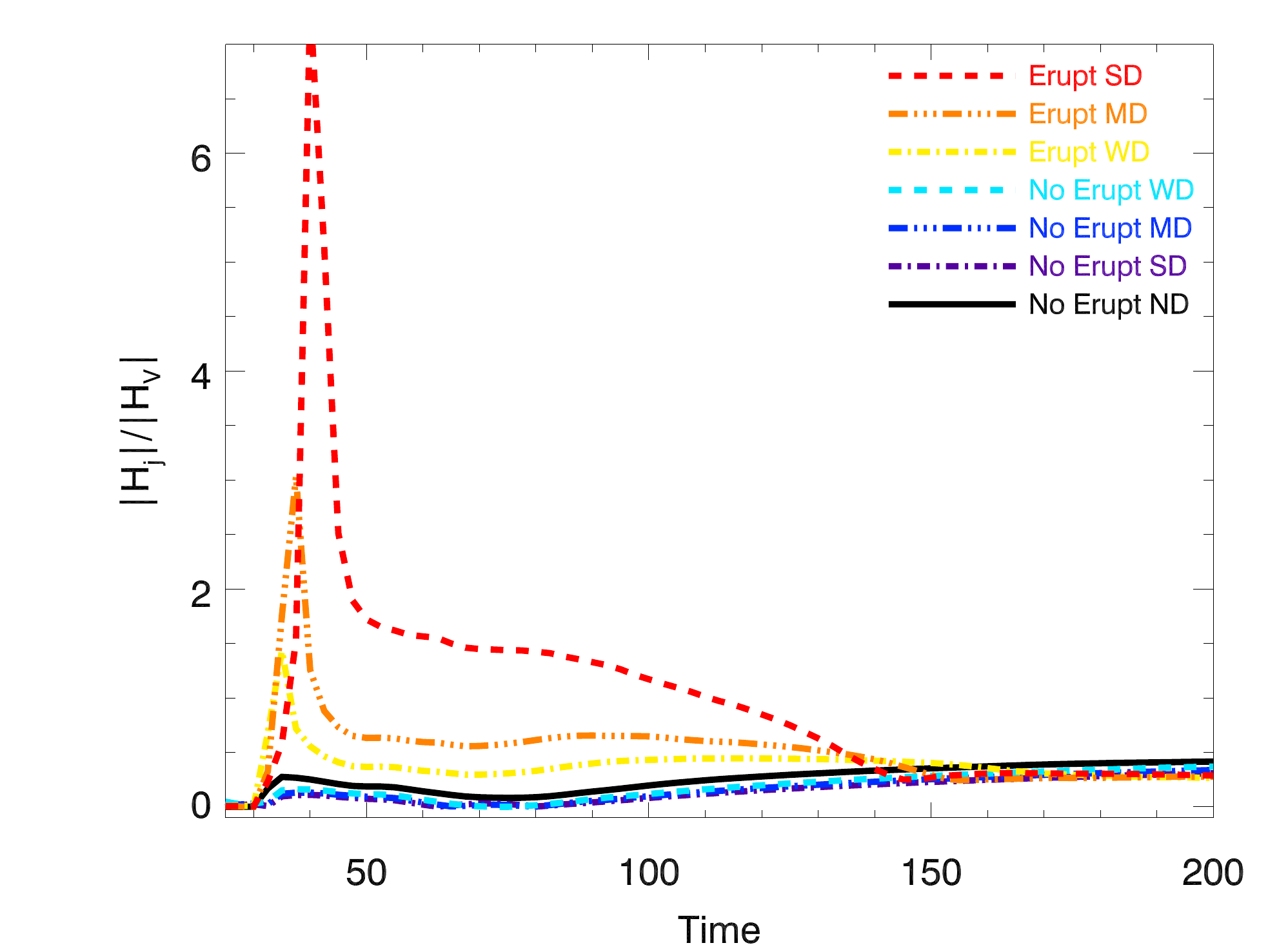}
  \caption{Time evolution of the helicity ratio $|\Hj|/|H_V|$  for the seven parametric simulations. The labels are similar to \fig{BFlux}. 
}
  \label{Fig:Hjfraction}
\end{figure}

\fig{Hjfraction} presents the ratio of $|\Hj|$ to the total helicity $|H_V|$ for the seven parametric simulations. Because these curves correspond to a ratio, and because $H_V$ is roughly null before $t\sim 30$ (magnetic flux only increases from that time on, we only plot values after that time in order to remove the spurious values resulting from the division by an infinitely small value. The non-eruptive curves are approximately constant throughout the simulation, with values that do not exceed $0.4$. For the four non-eruptive simulations, $\Hj$ always remains a minor contributor to $H_V$. 

On the contrary, the eruptive simulations all present high values of $|\Hj|/|H_V|$ during the first phase of the simulation. Immediately after the start of the emergence at $t \sim 30$, the curves present a very fast rise, with a peak between $t=35$ and $t=40$. The values of $|\Hj|/|H_V|$ even  exceed $1$, indicating an opposite sign between $\Hj$ and $\Hpj$. Helicity injection with opposite sign through the photospheric surface  has been reported in several observed cases of eruptive active regions \citep{Park12,Vemareddy12,Vemareddy16}. Here, the level of the peak appears to be correlated with the strength of the arcade. It should be noted however that the values of $H_V$ are still very small at these times, amounting to less than $2\%$ of the helicity eventually injected. After this initial peak, the values of $|\Hj|/|H_V|$ remain high, relative to the non eruptive simulations, with values globally above $0.45$. The level of $|\Hj|/|H_V|$ is also directly correlated with the strength of the arcade field, the stronger arcade  field presenting a larger ratio than the medium arcade. The lower arcade presents the smaller ratio among the eruptive simulations, although still markedly higher than the non-eruptive simulation, with values two to four times higher than the no-arcade case, and more than five times higher than with the other stable runs. The curves remain relatively constant for some time, eventually decreasing after $t\sim 85$ and finally, after $t\sim 135$ joining the group of the curves of the non-eruptive simulations.

The values $|\Hj|/|H_V|$ are not only higher for the eruptive simulations compared to the non-eruptive ones, but they are only so during the pre-eruptive and eruptive phase of the simulations. In the post-eruption phase of the eruptive simulations, when the system is not eruptive, these values are back to a low value, below $0.4$, typical of the non-eruptive simulations. Furthermore, during the eruptive phase, the ratio $|\Hj|/|H_V|$ is also markedly higher when the strength of the surrounding arcade is higher which, as noted by L14, is also related to a higher propensity for the system to erupt. Indeed, the higher arcade strength was associated with an earlier eruption of the system and a larger amount of reconnection. The ratio $|\Hj|/|H_V|$ thus appears as a very interesting criterion for qualifying and possibly quantifying the eruptivity of a system in solar-like conditions.

\section{Eruptivity criteria comparison} \label{s:Ecrit}

In the previous sections, several scalar quantities that can characterize the magnetic field have been computed and their evolution analyzed and compared for the seven parametric simulations. Their ability to constitute a pertinent criterion of the eruptivity of the system has been qualitatively discussed. Only positive-defined quantities are being considered in the present study and constructed so that eruptivity could possibly be associated with a high value of these proxies. Let us note that such positive criteria can always be built by the use of opposite or the inverse and modulus functions. 
In order to quantify the quality of a good eruptivity proxy, we compute different parameters that evaluate their distribution between the different simulations: the mean value, $\mu$, and the relative standard deviation, $C_v$, of a given quantity at a given time for all seven simulations; the mean values, $\mu_{\rm Erupt}$ and $\mu_{\rm No\ Erupt}$ only considering respectively the eruptive/non-eruptive simulations, and their ratio $\eta=  \mu_{\rm Erupt}/\mu_{\rm No\ Erupt}$. 

A quantity which is not able to distinguish between the different simulations will have $C_v$ close to 0, as well as a value of $\eta$ close to 1. This quantity thus does not possess the quality of a good proxy. 

A high value of $C_v$ indicates that this quantity is discriminating between the different simulations, although not necessarily their eruptive/non-eruptive character. Associated with a value of $\eta$ close to 1, this means that this proxy is mostly sensitive to the strength of the surrounding arcade rather than the eruptivity. 

A small value of $\eta$, close to 0, indicates that this quantity is significantly higher for the non-eruptive simulations. This generally means that this quantity is not a good proxy for eruptivity. On the contrary, a high value of $\eta$, associated with a large value of $C_v$, is what is required from a good eruptivity proxy since it indicates that this quantity tends to be significantly higher for the eruptive simulations compared to the non-eruptive ones. For the eruptive simulations, a good eruptivity proxy should only have high values of $C_v$ and $\eta$ during the pre-eruptive and the eruptive phase and not during the post-eruptive phase. 


\begin{figure}[ht]
  \setlength{\imsize}{0.49\textwidth}
  \includegraphics[width=\imsize,clip=true]{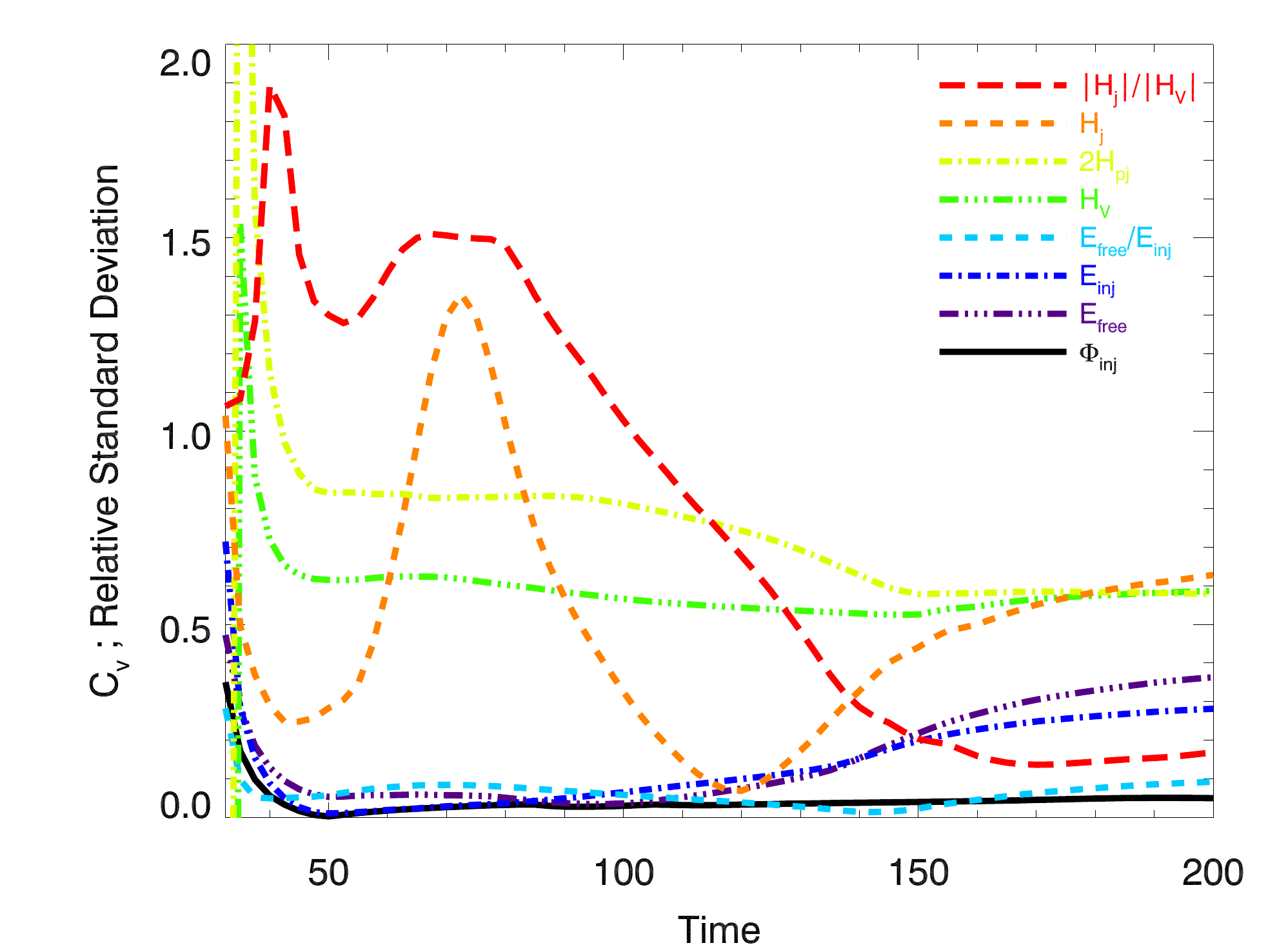}
  \includegraphics[width=\imsize,clip=true]{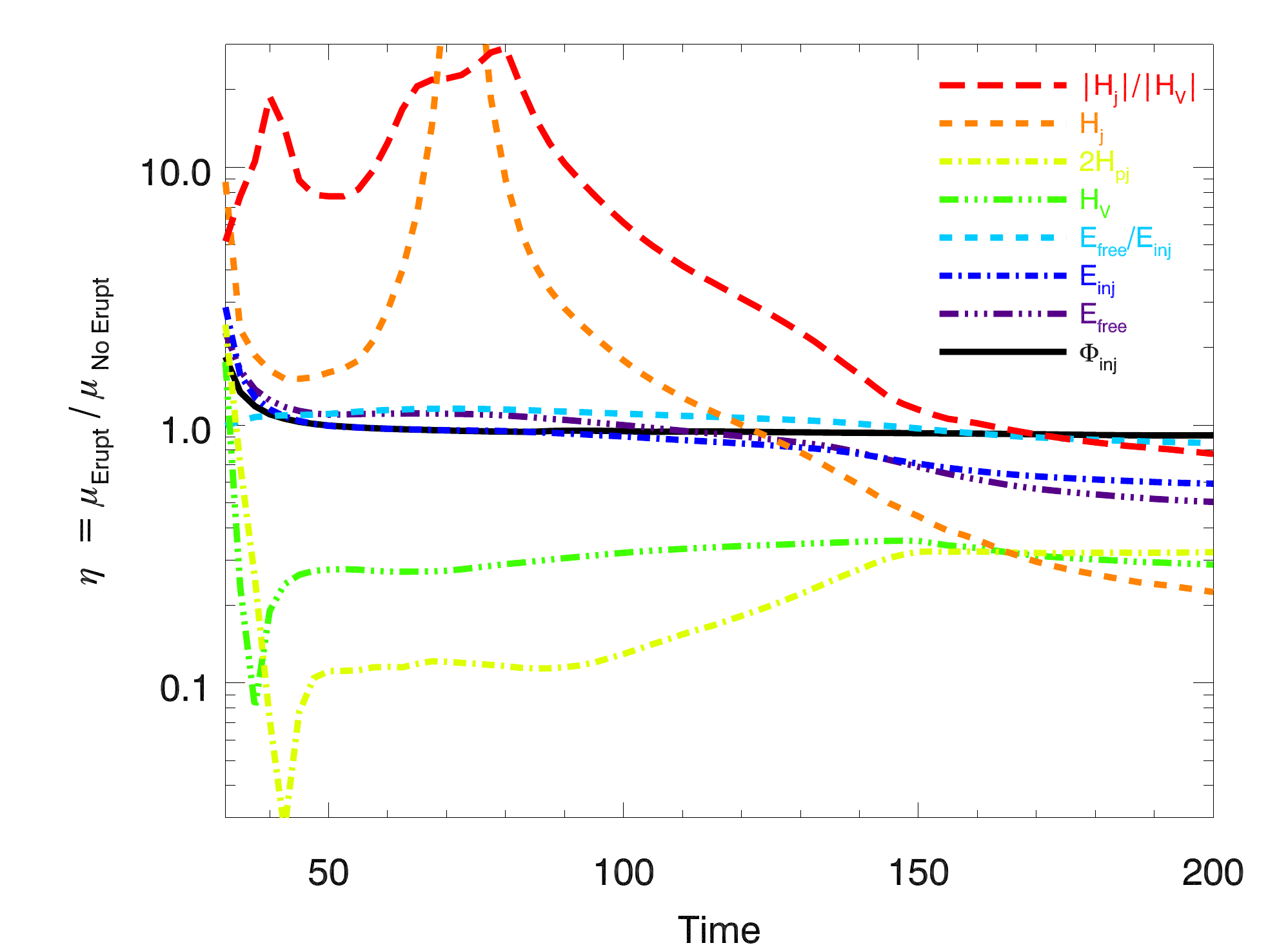}
  \caption{Time evolution of the relative standard deviation, ($C_v$, top panel) and ratios of the mean-eruptive and non-eruptive values ($\eta$, bottom panel) of several potential criteria for eruptivity. The [black continuous, purple three-dot-dashed, blue dot-dashed, cyan dashed, green three-dot-dashed, yellow dot-dashed, orange dashed, red long-dashed] curve corresponds, respectively, to the quantity [$\Phi_{inj}$, $E_{free}$, $E_{inj}$, $E_{free}/E_{inj}$, $H_{V}$, $2\Hpj$, $\Hj$, $|\Hj|/|\Hpj$].
}
  \label{Fig:Crit}
\end{figure}

\fig{Crit} presents the evolution of $\eta$ and of the relative standard deviation between the different simulations. The curves of $C_v$ show that the criteria based on magnetic flux and magnetic energy, namely, $\Phi_{inj}$, $E_{inj}$, $E_{free}$, $E_{free}/E_{mag}$, $E_{free}/E_{inj}$ are not able to distinguish between the different simulations. Their relative standard deviation remains very small, $<20\%$, during the pre-eruptive phase, for $t\in[40,70]$. On the contrary the magnetic-helicity based proxies, $H_V$, $\Hj$, $\Hpj$, $|\Hj|/|H_V|$ present high standard deviations indicating that the different simulations are strongly discriminating between these quantities. $|\Hj|/|H_V|$ possesses the noticeable property of being very high in the pre-eruptive phase and in the eruptive phase, while decreasing in the post-eruptive phase.

The curves of $\eta$ confirm again that the flux injection, $\Phi_{inj}$, is extremely similar for the eruptive and the non-eruptive simulations. The mean value of the eruptive simulations is constantly almost equal to the mean value of the non-eruptive ones. This quantity thus does not present any quality looked for in an eruptive proxy. The same is true for $E_{inj}$. The total volume helicity $H_V$ and $\Hpj$, while displaying distinctive behavior between eruptive and non-eruptive, have very low values of $\eta$. This indicates that the eruptive simulation tends to have weaker values than the non-eruptive one. We observed in \sect{H}, that the highest values of these quantities were eventually reached by the non-eruptive simulations during the post-eruption phase. It is therefore not possible to define a threshold from these quantities, which are thus poor eruptivity criteria.

The free magnetic energy, as well as $E_{free}/E_{mag}$ and  $E_{free}/E_{inj}$, shows a weak tendency to be higher for the eruptive simulation during their eruptive phase. $E_{free}$ nonetheless has an $\eta$ value that drops markedly below 1 in the post-eruptive phase. As shown in \fig{E}, the non-eruptive simulations present the highest values of  $E_{free}$ during that period. No threshold on $E_{free}$ can thus be built in the present simulation framework. While being notably higher than $1$ during the pre-eruptive phase, $\eta(E_{free}/E_{inj})$ decreases in the post-eruptive phase and becomes close to 1. Nonetheless, the value of $\eta$, of approximately$1.15$ during the pre-eruptive phase, is not very high and thus may be of little practical use with real data. In addition, as already noted in \sect{E}, the non-eruptive simulations reach greater values of $E_{free}/E_{inj}$ in the post-eruptive phase, even though no eruption is present (cf. L13). The ratio $E_{free}/E_{inj}$ therefore does not likely constitute a reliable proxy for eruption prediction.

The current magnetic helicity $\Hj$ presents a high value of $\eta$ during the pre-eruptive phase, but then presents a low value during the post-eruptive phase, since the non-eruptive simulations present the highest values of $\Hj$.  Similarly to $E_{free}$, no threshold on $\Hj$ can be constructed that would enable the prediction of the eruptivity of our parametric simulation. 

Finally, \fig{Crit} confirms that the ratio $|\Hj|/|H_V|$ is an extremely efficient proxy of eruptivity for the simulations. This quantity presents clear variations of $C_v$ from the pre-eruption phase to the post-eruption phase, and also presents a very high $\eta$, with values $>5$, in the pre-eruptive phase (for $t\in[30,120]$).  The values of $\eta(|\Hj|/|H_V|)$ decrease during the eruptive phase and eventually  become close to $1$ during the non-eruptive phase, for $t>150$, indicating that the eruptive and non-eruptive simulations are no longer distinguishable. This is expected since none of the simulations present any eruptive behavior during that last phase. An eruptivity threshold can easily be built from the ratio of  $|\Hj|/|H_V|$. This quantity thus possesses a very strong potential to allow the prediction of solar eruptions.
  
\section{Conclusion and discussion} \label{s:Conclusion}

In the present study, we have computed and compared the magnetic energy and helicity evolution of the coronal domain of seven parametric 3D MHD simulations of flux emergence, initially presented in 
\citet{Leake13b} and \citet{Leake14a}. These numerical experiments, while only modifying a unique parameter - the strength and direction of the background coronal field - led either to a stable configuration or to an eruptive behavior with the ejection of a CME-like magnetic structure. These simulations represent a particularly interesting dataset that enables us to search for eruptivity criteria.

Following the method of decomposition of the magnetic energy of \citet{Valori13} and the method for computing relative magnetic helicity presented in \citet{Valori12}, we have computed different magnetic flux-, energy-, and helicity-based quantities throughout the evolution of the systems.  As expected from the numerical setup, we noted that all the simulations presented a quasi-similar injection of magnetic flux. We have found that unlike magnetic flux and energy, relative magnetic helicity very clearly discriminates between eruptive and stable simulations.

We have however found that the total magnetic helicity was not correlated with a stronger eruptive behavior. Non-eruptive simulations, in fact, presented a higher absolute value of the total magnetic helicity compared to the eruptive ones. Using a toy model we have shown that the non-eruptive simulation possessed self helicity of the emerging flux rope of the same sign as the mutual helicity between the emerging flux rope and the coronal background field. Eruptive simulations presented a lower total helicity because the self and mutual helicities were of opposite sign. 

Our results thus confirm those from \citet{Phillips05}, stating that the total magnetic helicity is not a determining factor for CME initiation and that there might not be a universal threshold on eruptivity based on total magnetic helicity. We however argue against their conclusion that helicity in general is unimportant. Their setup, similarly to our eruptive cases, presents large helicities of opposite signs. The decomposition of helicity, if not its distribution, seems to be related to enhanced eruptive behavior. Using the helicity decomposition of the relative magnetic helicity in the current-carrying magnetic helicity, $\Hj$ and its counterpart $2\Hpj$, introduced by \citet{Berger03}, eruptive and non-eruptive cases present noticeably distinct behavior, with the eruptive simulations presenting significantly greater values of $\Hj$ during the pre-eruptive phase.

Comparing the different quantities in their capacity to efficiently describe the eruptivity status of the different simulations during their evolution, we noted that while the ratio of the free magnetic energy to the injected magnetic energy, $E_{free}/E_{inj}$ is higher for the eruptive simulations, in their pre-eruptive phase only, it presents two drawbacks: the values of this quantity are only marginally higher (by a few $\%$) for the eruptive simulation, compared to the non-eruptive ones, and the non-eruptive simulations can reach values of a similar amplitude to the eruptive case. The definition of an eruptivity threshold seems to be difficult to determine with such a quantity.

The quantity that appears as an excellent eruptivity proxy is the ratio of the current carrying magnetic helicity to the total helicity, $|\Hj|/|H_V|$. This ratio is several times higher for the eruptive simulations compared to the non-eruptive simulations during the pre-eruptive phase, and similar for all simulations during the non-eruptive phase. None of  the non-eruptive simulations present values of $|\Hj|/|H_V|$ higher than $0.45$ while all the eruptive simulations can reach values three times that threshold. The strong and medium arcade cases, which present the fastest eruptive behaviors, have values of $|\Hj|/|H_V|$ higher than $0.45$ for almost all their pre-eruptive phase. In the framework of the present parametric simulations, the quantity $|\Hj|/|H_V|$ thus constitutes an excellent eruptivity proxy.
 
 The use of parametric numerical simulations has enabled us to highlight the existence of a very promising proxy of eruptivity that could allow for improvement in our forecast of solar eruptions. One should however be conscious of the limits of our approach. The first limitation is of course the level of realism of the present simulations compared to the real Sun. In addition to the inherent limits of the MHD paradigm, two aspects of the numerical setup may deviate from the conditions found in the Sun. The first is the relative intensity of the background field to the emerging flux rope; the magnetic flux of the background coronal arcade is of similar scale to the flux injected by the emerging flux rope. This is different from most large-scale active regions for which the initial field is notably smaller than the emerging flux. While non-dimensional, the scale of these simulations corresponds to a small-scale emerging structure.  In addition, the amount of helicity injected by the emerging structure is relatively high compared to typical solar values. Observations have shown that emerging active regions present normalized helicities $\Hn$ (cf \eq{Hnorm}) comprised between $0.01$ and $0.1$ \citep{Labonte07,Jeong07,TianL08,YangS09,Demoulin09} which is several times smaller than the values obtained here. The emerging structure in the present simulation is thus injecting more twist into the coronal domain than typical active regions do. These limitations do not question the validity of our study and its results, but only its uncritical application to real data.

Furthermore, unlike most observational studies that, when testing eruptivity proxies, are incorporating active regions of various size, with magnetic fluxes ranging over orders of magnitude, our present dataset is limited to simulated active regions of quasi-identical size. Observational statistical studies show that larger active regions (in the sense of magnetic flux) have a greater eruptivity and flare productivity \citep[e.g.,][]{Labonte07,ParkSH10,Tziotziou12}. The present study is thus not addressing this property. The eruptivity proxy $|\Hj|/|H_V|$ may only be a determinant eruptivity criterion between active regions of similar magnetic flux. When comparing regions of different size, its ability may be reduced. Most observational studies indeed show that it is a combination of proxies that gives the best eruptivity criterion \citep{Bobra15,Bobra16}. It is however interesting to note that $|\Hj|/|H_V|$ is an intensive quantity and thus its capacity to predict eruptions may be independent from the active region size.
 
Additionally, our results are so far limited to the modeling framework of the flux emergence simulations of \citet{Leake13b,Leake14a}. Other flux emergence simulations, with different setups (e.g., horizontal coronal field) should also be tested. Completely different models, such as line-tied coronal eruptive simulations, for example, should also be studied. This is the goal of an ongoing study (Zuccarello et al., in preparation) that focuses on the analysis of the helicity content of the parametric simulations of \citet{Zuccarello15}.

The final caveats that may limit the usage of $|\Hj|/|H_V|$ are intrinsically related to our limited understanding of the properties of magnetic helicity. Relative magnetic helicity as defined by \eq{Hrel}, is not a simply additive quantity \citep{Berger00}. While all the computations done in the present study have been performed using the very same coronal domain, it is not guaranteed that the values of the ratio $|\Hj|/|H_V|$ do not depend on the volume and boundaries of the  domain being studied. It is important to keep this in mind, since it implies that the value of $0.45$ discussed previously is likely to only be valid with regards to the present simulations given the size of the coronal domain chosen here. Further investigation is needed in order to better comprehend the properties of magnetic helicities.

Despite these limitations, our study shows that magnetic-helicity-based quantities may be crucial in predicting solar eruptions. Because of its difficult estimation  in observational cases, this class of quantity has been largely neglected in systematic studies searching for eruptivity proxies. Based on the present work, we argue that an additional effort should be carried out to test relative magnetic-helicity-based quantities against eruptive behavior of observed active regions. We have demonstrated that the ratio $|\Hj|/|H_V|$ appears to be a very promising intensive quantity. 

The theoretical reasons explaining why this ratio is so efficient at describing the eruptive states of our parametric simulations still need to be fully explored. It is however interesting to note that $\Hj$ corresponds to a description of the current part of the field while the $|H_V|$, which is dominated by $2\Hpj$, includes the relative properties of the current-carrying field in relation to its surrounding field. In the framework of the torus instability \citep{Kliem06,Demoulin10,Zuccarello15}, the criterion of instability is also related to the relative properties of the current-carrying structure with regards to its surrounding confining field. Here we speculate that the ratio $|\Hj|/|H_V|$ may somehow be related to the instability criterion of the torus instability. Further theoretical research should be carried out to further understand the link between the relative magnetic helicity decomposition and the eruptivity of a system.

\begin{acknowledgements}
The authors would like to thank Sung-Hong Park and an anonymous referee for their remarks that helped to improve the quality of this manuscript.
EP acknowledge the support of the French Agence Nationale pour la Recherche  through the HELISOL project, contract n$^\circ$ ANR-15-CE31-0001, as well as the FLARECAST project, funded by the European Union's Horizon2020 research and innovation programme under grant agreement n$^\circ$ 640216.
 JEL and MGL acknowledge support from the Chief of Naval Research and from the NASA Living with a Star program. The flux emergence simulations analyzed in this work were performed under a grant of computer time from the US Department of Defense High Performance Computing program. GV acknowledges the support of the Leverhulme Trust Research Project Grant 2014-051. FPZ. is a Fonds Wetenschappelijk Onderzoek (FWO) research fellow (Project No. 1272714N). KD acknowledges support from the Computational and Information Systems Laboratory and the High Altitude Observatory of the National Center for Atmospheric Research, which is sponsored by the National Science Foundation. GV and EP thanks ISSI, its members and the participants of the ISSI international team on {\it Magnetic Helicity estimations in models and observations of the solar magnetic field} where this work has been discussed and commented. 
\end{acknowledgements}

 \bibliographystyle{aa}  
\bibliography{HeliErupt}       
\IfFileExists{\jobname.bbl}{}  
{ 
\typeout{} 
\typeout{****************************************************} 
\typeout{****************************************************} 
\typeout{** Please run "bibtex \jobname" to obtain}  
\typeout{**the bibliography and then re-run LaTeX}  
\typeout{** twice to fix the references!} 
\typeout{****************************************************} 
\typeout{****************************************************} 
\typeout{} 
 }

\appendix

\section{Gauge invariance} \label{s:Ginv}

The helicity computation method of \citet{Valori12} allows the estimation of helicity quantities using different sets of gauges. The potential vectors are computed using the DeVore gauge \citep{DeVore00}, that is, $A_z=0$. As discussed in Section 3.2 of \citet{Pariat15b}, the derivation is performed by choosing a reference plane $z_i$, here either the top or bottom boundary of our dataset. At this boundary, \citet{Valori12} proposed either choosing an integral or a derivative methodology. In the latter case, noted "DV-C" hereafter, $\vAp$, is simultaneously fulfilling the DeVore and Coulomb ($\divAp=0$) gauge conditions. The former case is simply noted "DV". Assuming $A_z(z_i)=A_{p,z}(z_i)$, we can therefore derive \eq{Hrel} using four different sets of gauges, noted "DV bot", "DV top", "DV-C bot", and "DV-C top", with "top" and "bot" referring to the choice of $z_i$ as respectively the top or bottom boundary.

\begin{figure}[ht]
  \setlength{\imsize}{0.49\textwidth}
   \includegraphics[width=\imsize,clip=true]{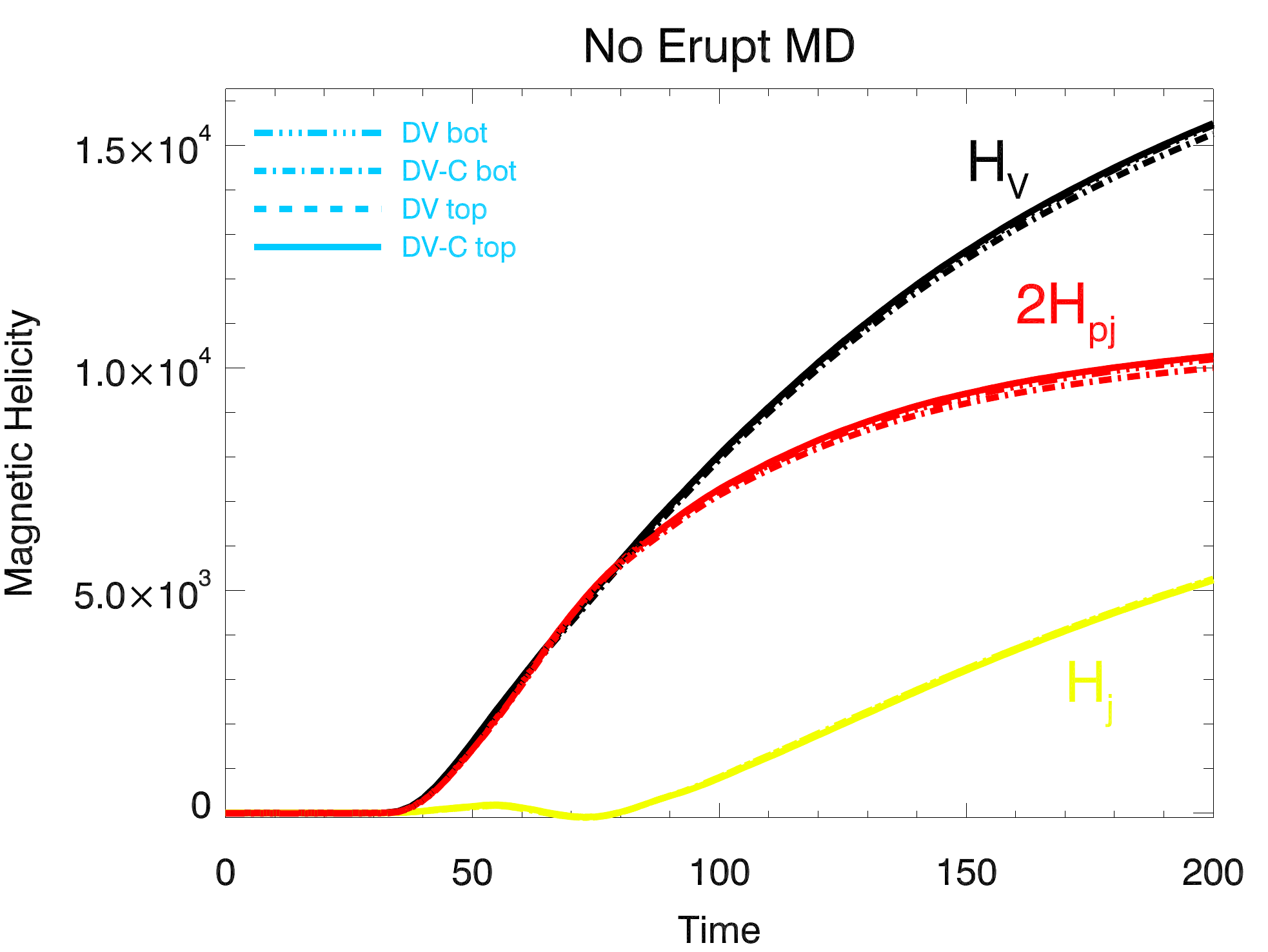}
  \includegraphics[width=\imsize,clip=true]{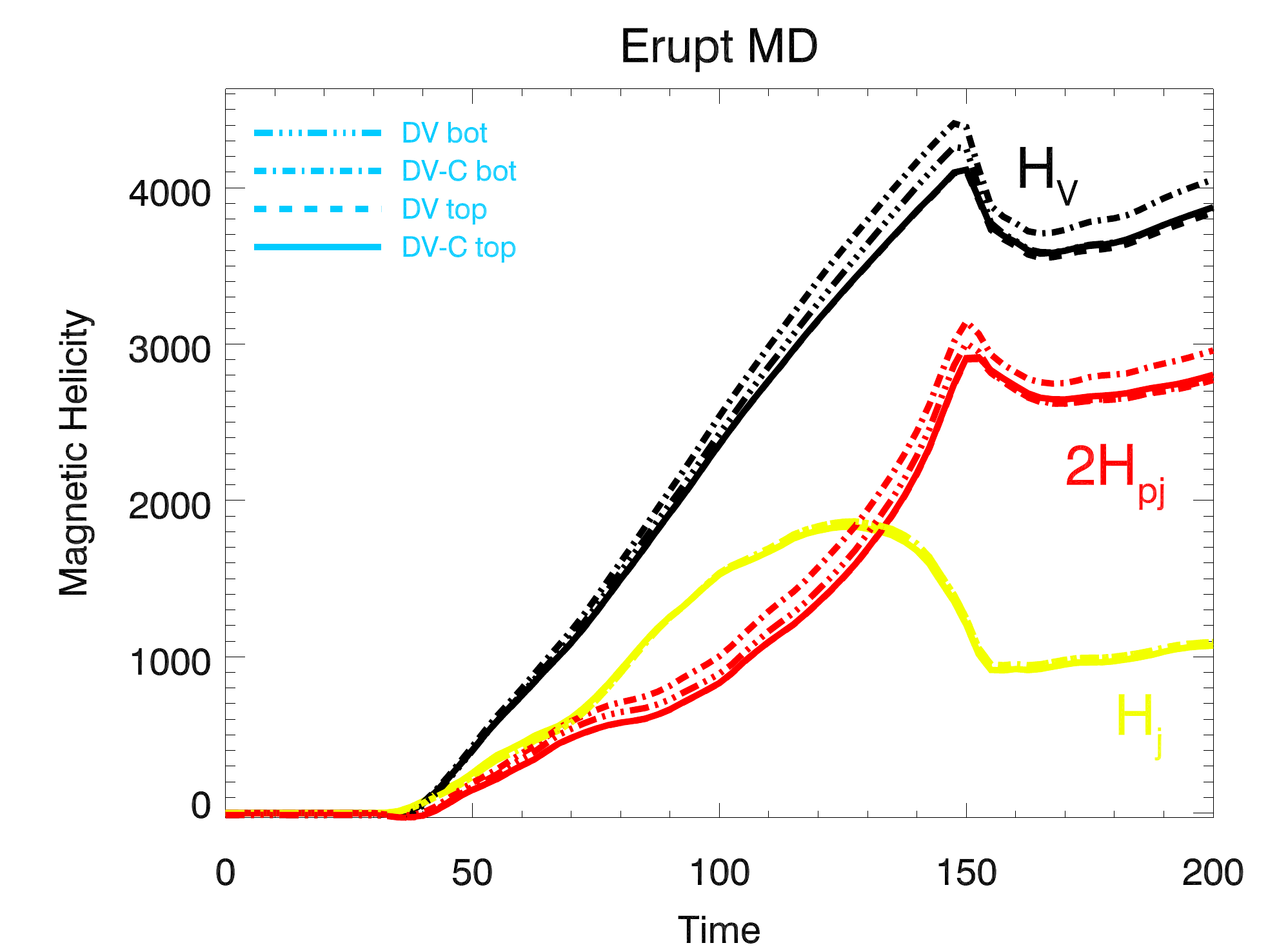}
  \caption{Helicity and its decomposition in the medium arcade simulations for the non-eruptive (top panel) and eruptive (bottom panel) cases. The [black, red, yellow] curves corresponding to [$H_V$,$2\Hpj$,$\Hj$], respectively. The helicities are computed using  the [DV bot, DV top, DV-C bot, DV-C top] gauge sets,  corresponding to the [three-dot-dashed, dashed, dot-dashed, continuous] curves, respectively.
}
  \label{Fig:HjHpjgauges}
\end{figure}

\fig{HjHpjgauges} displays the computation of $H_V$, $\Hj$ , and $\Hpj$ in these four different gauges for both the eruptive and non-eruptive cases with the medium-strength arcade (MD cases). One can remark that $H_V$, $\Hj$ , and $\Hpj$ are gauge invariant. For the non-eruptive case, the maximum relative standard deviation of the values of $H_V$ is only $1\%$ while it is slightly higher, $4\%$, for the eruptive simulations. The eruptive simulations indeed represent a more arduous case for helicity since both the bottom and the top boundary are significantly evolving in time. One observes that $\Hj$ is particularly gauge invariant with maximum standard deviation of $2\%$ for both simulations. $\Hpj$ contains most of the gauge dependance of $H_V$ with relative standard deviations which amount to $2\%$ and $10\%,$ respectively, for the non-eruptive and the eruptive cases. The other simulation cases present similar trends of gauge dependance.   
  
Overall, the distribution of values computed with different gauges provides us with rough estimates of the precision of the helicity computation. We can thus estimate that $H_V$ has an average $5\%$ measurement error. Any difference above this value can thus be considered as significant, which is the case of all the comparisons between simulations that are done in this paper. One can note that the gauges computed starting at the top boundary tend to be more consistent with each other, which is likely a result of the distribution of solenoidal errors specific to this simulation.  As already noted in \citet{Valori12} and \citet{Pariat15b}, the DeVore-Coulomb approach is also less prone to numerical error. In the main text of this study, the helicity values are obtained using the "DV-C top" gauge set.

\section{Magnetic helicity conservation} \label{s:Hcons}

\begin{figure}[ht]
  \setlength{\imsize}{0.49\textwidth}
   \includegraphics[width=\imsize,clip=true]{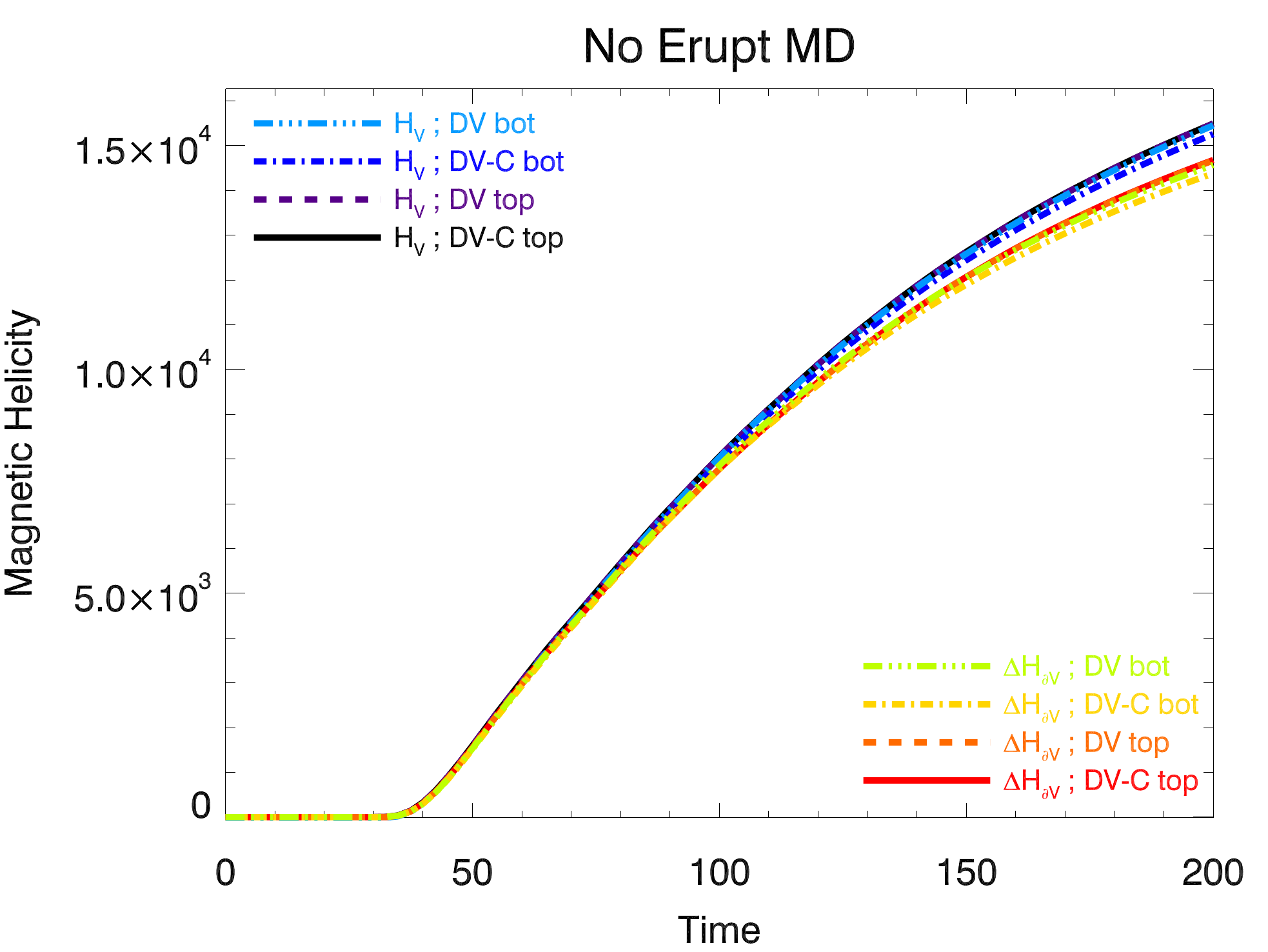}
  \includegraphics[width=\imsize,clip=true]{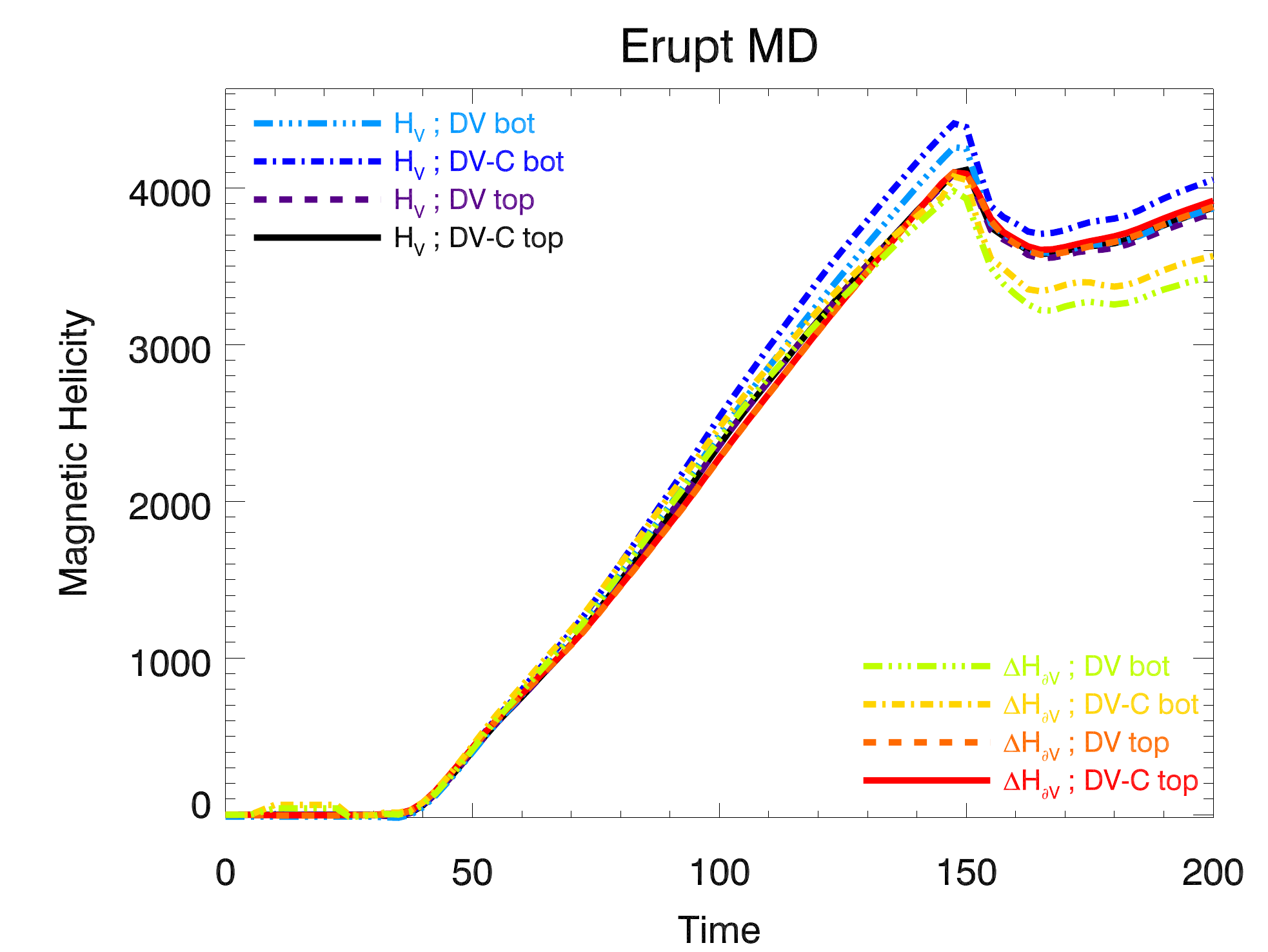}
  \caption{Helicity variation and accumulated flux in the medium  arcade simulations for the non-eruptive (left panel) and eruptive (right panel) cases. The volume helicity estimations, $H_V$, are plotted in black, purple, blue, and cyan while the time-integrated flux of helicity, $\Delta H_\surf$, are plotted in red, orange, yellow, and green. The helicities are computed using  the [DV bot, DV top, DV-C bot, DV-C top] gauge sets,  corresponding to the [three-dot-dashed, dashed, dot-dashed, continous] curves, respectively. }
  \label{Fig:Hgauges}
\end{figure}

In addition to the volume helicity, the helicity computation method of \citet{Valori12} allows for  the derivation of the complete time integrated flux of helicity through, for example, the six boundaries of the cartesian domain \citep[cf.][]{Pariat15b}. This enables the determination of the helicity, $\Delta H_\surf$, accumulating in the domain in time:
\BE 
\Delta H_\surf (t)  = \int_0^t  \Ftot(\tau) ~\rmd \tau  \label{eq:defHsurftot} \\
,\EE
with $\Ftot$ the flux of helicity given by equation (30) of \citet{Pariat15b}. The computation of $\Delta H_\surf$ requires both the knowledge of the velocity and magnetic field distribution on $\surf$. It is independent of the derivation of $H_V$. $\Delta H_\surf$ is theoretically gauge invariant \citep{Pariat15b}. \fig{Hgauges} presents the measurements of $\Delta H_\surf$ with the four gauge sets considered in \app{Ginv}, for the medium arcade   cases. Our results show indeed that $\Delta H_\surf$ is relatively gauge invariant, with a $1\%$ maximum relative standard deviation of the values for the non-eruptive simulation. Here again the eruptive case shows slightly larger error, with a $6\%$ relative standard deviation. As already noted in \citet{Pariat15b}, because the computation of $\Delta H_\surf$ requires more integration, and in particular a time integration, it tends to be more sensitive to numerical errors than $H_V$ and therefore presents a slightly stronger gauge dependance. 
The gauge sets computed from the top boundary are also more consistent, being less sensitive to errors.   

In ideal MHD, magnetic helicity is a conserved quantity \citep{Woltjer58}. Even when non-ideal processes, such as magnetic reconnection, are developing, magnetic helicity is believed to be a quasi-conserved quantity \citep{Taylor74,Berger84}. In a recent numerical simulation of a solar-like active event, with intense magnetic reconnection, \citet{Pariat15b} show that magnetic helicity is extremely well conserved, with a $<2\%$ dissipation, particularly with regards to the magnetic energy dissipation which was more than 30 times higher. Following \cite{YangS13} and \citet{Pariat15b}, the comparison of the time-integrated helicity flux and volume helicity enables one  to study the helicity conservation in the numerical simulation, and how close the numerical scheme is to ideal MHD behavior. In the present simulations, since the helicity in the system is initially null, assuming perfect helicity conservation, $\Delta H_\surf$ and $H_V$ should be equal. 
For the medium  arcade cases, one observes on \fig{Hgauges} that, independently of the gauge set used, the curves of $\Delta H_\surf$ and of $H_V$ match well. This indicates a good level of conservation of magnetic helicity in these simulations. Using the relative accumulated helicity difference criteria, $\epsH$, defined in equation (43) of \citet{Pariat15b}, we note maximum values, using the least-conservative gauge set, of $6\%$ for the non-eruptive simulation and $15\%$ for the eruptive one. This demonstrates that while not perfect, the code is able to conserve magnetic helicity relatively well. Similar helicity-conservation trends are obtained for all the simulation datasets used in this study.

\end{document}